\documentclass[prd,aps,superscriptaddress,preprintnumbers,eqsecnum,nofootinbib,notitlepage]{revtex4-2}
\usepackage[utf8]{inputenc}
\usepackage{amsfonts,amsmath,amssymb,bm,subfigure,graphicx,ulem}
\usepackage{multirow,hyperref}
\usepackage{xcolor}

\newcommand{\nn}{\nonumber \\}
\newcommand{\D}{{\rm d}}

\newcommand{\g}{g}
\newcommand{\Mpl}{M_*}
\newcommand{\gf}{g_{\rm eff}}
\newcommand{\eff}{{\rm eff}}
\newcommand{\cutoff}{\Lambda_*}

\newcommand{\xx}{\boldsymbol{x}}

\newcommand{\spatialR}{ {}^{(3)}\!R }

\newcommand{\spatialRST}{ {}^{(3)}\!R }

\begin{document}
\baselineskip=12pt

\preprint{YITP-24-56, RUP-24-8, IPMU24-0016}
\title{CMB spectrum in unified EFT of dark energy:\\ scalar-tensor and vector-tensor theories
}
\author{Katsuki Aoki}
\affiliation{Center for Gravitational Physics and Quantum Information, Yukawa Institute for Theoretical Physics, Kyoto University, 606-8502, Kyoto, Japan}

\author{Mohammad Ali Gorji}
\affiliation{Departament de F\'{i}sica Qu\`{a}ntica i Astrof\'{i}sica, Institut de Ci\`{e}ncies del Cosmos, Universitat de Barcelona, Mart\'{i} i Franqu\`{e}s 1, 08028 Barcelona, Spain}

\author{Takashi Hiramatsu}
\affiliation{Department of Physics, Rikkyo University, Toshima, Tokyo 171-8501, Japan
}

\author{Shinji Mukohyama}
\affiliation{Center for Gravitational Physics and Quantum Information, Yukawa Institute for Theoretical Physics, Kyoto University, 606-8502, Kyoto, Japan}
\affiliation{Kavli Institute for the Physics and Mathematics of the Universe (WPI), The University of Tokyo, Kashiwa, Chiba 277-8583, Japan}

\author{Masroor C. Pookkillath}
\affiliation{Centre for Theoretical Physics and Natural Philosophy, Mahidol University, Nakhonsawan Campus,  Phayuha Khiri, Nakhonsawan 60130, Thailand}

\author{Kazufumi Takahashi}
\affiliation{Center for Gravitational Physics and Quantum Information, Yukawa Institute for Theoretical Physics, Kyoto University, 606-8502, Kyoto, Japan}

\date{\today}

\begin{abstract}
We study the cosmic microwave background (CMB) radiation in the unified description of the effective field theory (EFT) of dark energy that accommodates both scalar-tensor and vector-tensor theories. The boundaries of different classes of theories are universally parameterised by a new EFT parameter $\alpha_V$ characterising the vectorial nature of dark energy and a set of consistency relations associated with the global/local shift symmetry. After implementing the equations of motion in a Boltzmann code, as a demonstration, we compute the CMB power spectrum based on the $w$CDM background with the EFT parameterisation of perturbations and a concrete Horndeski/generalised Proca theory. We show that the vectorial nature generically prevents modifications of gravity in the CMB spectrum. On the other hand, while the shift symmetry is less significant in the perturbation equations unless the background is close to the $\Lambda$CDM, it requires that the effective equation of state of dark energy is in the phantom region $w_{\rm DE}<-1$. The latter is particularly interesting in light of the latest result of the DESI+CMB combination as the observational verification of $w_{\rm DE}>-1$ can rule out shift-symmetric theories including vector-tensor theories in one shot.

\end{abstract}

\maketitle

\section{Introduction}

The study of our Universe has achieved high maturity in terms of techniques and approaches to address the crucial question: ``How does our Universe work?'' Theoretically, it has given birth to the modern gravitational theory known as general relativity (GR). On the observational side, it is realised that our Universe is expanding. Later, more surprises like the discovery of cosmic microwave background (CMB) radiation~\cite{Penzias:1965wn} and the late-time cosmic acceleration~\cite{SupernovaSearchTeam:1998fmf, SupernovaCosmologyProject:1998vns} have been achieved by means of the missions like WMAP~\footnote{https://map.gsfc.nasa.gov/}, PLANCK~\footnote{https://www.cosmos.esa.int/web/planck}, and large-scale surveys. The modern cosmology is then constructed based on these theoretical and observational discoveries. 

As a result of the developments in the cosmological surveys, it is evident that the Universe today is dominated by unknown mysterious components, viz, dark matter and dark energy (DE) if GR is the correct description of gravity at cosmological scales. 
In particular, in cosmology based on GR the DE component is an unavoidable component that acts as an agent for the late-time cosmic acceleration. Moreover, recent developments in observations hint at differences in the Hubble parameter $H_0$ \cite{Riess:2021jrx, DiValentino:2021izs} and the amplitude of the density fluctuation $S_{8}$ \cite{DiValentino:2020vvd} measured by CMB observations and the local/large-scale structure surveys.

Instead of introducing such an unknown energy component, one can propose modified theories of gravity. Modified gravity is also one of the promising directions to resolve the observational tensions~\cite{Perivolaropoulos:2021jda}. For example, there have been many modified gravity models addressing the inevitable Hubble tension~\cite{DiValentino:2021izs,DeFelice:2020sdq, DeFelice:2020cpt} (see also~\cite{Karwal:2016vyq, Poulin:2018cxd, Poulin:2023lkg} for the early DE scenarios). Most of the modified gravity theories introduce additional degree(s) of freedom~\cite{Clifton:2011jh,Koyama:2015vza,Ferreira:2019xrr,Arai:2022ilw}. The simplest case is to introduce a minimally coupled canonical scalar field, which is known as quintessence~\cite{Tsujikawa:2013fta}. Another possibility is to introduce a vector field~\cite{Golovnev:2008cf, Tasinato:2014eka}. Several generalisations of these simple pictures of scalar-tensor theories~\cite{Horndeski:1974wa, Nicolis:2008in,Deffayet:2013lga,Gleyzes:2014qga,Langlois:2015cwa,Crisostomi:2016czh,BenAchour:2016fzp,Takahashi:2022mew,Takahashi:2023jro,Takahashi:2023vva} and vector-tensor theories~\cite{Heisenberg:2014rta,Allys:2015sht,BeltranJimenez:2016rff,DeFelice:2016yws,Heisenberg:2016eld,Allys:2016jaq,Kimura:2016rzw,deRham:2020yet} have been suggested. To recognise the advantage of these various generalisations, one needs to study the cosmological perturbation of the theory, which gives new features that can be tested with cosmological observations. 

The effective field theory (EFT) approach has a great advantage, as the EFT can accommodate various phenomenological aspects of different modified gravity theories in a unified framework~\cite{Gubitosi:2012hu, Frusciante:2019xia}. In the EFT of DE, phenomenological effects exhibited in different theories are parameterised by time-dependent EFT functions in a universal way.
However, the conventional EFT of DE is an EFT of scalar-tensor theories and cannot make room for theories beyond scalar-tensor theories. On the other hand, there are theories of modified gravity with massive vector fields, known as Proca theory and its generalisations~\cite{Heisenberg:2014rta}. It was shown that generalised Proca theories have interesting phenomenology and can even have the potential to address the Hubble tension~\cite{DeFelice:2020sdq}.\footnote{However, it has to be noticed that generalised Proca theory at the late time cannot completely fix the Hubble tension.} It is therefore desirable to incorporate the vector-tensor theories on an equal footing with the scalar-tensor theories to understand their phenomenological relevance systematically. Some of the authors achieved this by formulating the EFT of vector-tensor theories~\cite{Aoki:2021wew}. It is not just the EFT of vector-tensor theories but serves as a unified description of scalar-tensor and vector-tensor theories as it recovers the EFT of scalar-tensor theories in an appropriate limit. 

The vector-tensor extension requires an additional EFT function~$\alpha_{V}$, which parameterises the strength of the vectorial nature of DE. On top of that, vector-tensor theories are viewed as gauging shift-symmetric scalar-tensor theories. The presence of shift symmetry, either global or local, is characterised by a set of consistency relations among some of the EFT functions and the Hubble expansion rate. In this work, we explore the impact of the new parameter~$\alpha_{V}$ and the consistency relations in the CMB power spectrum using a Boltzmann solver developed by one of the authors in Refs.~\cite{Hiramatsu:2020fcd,Hiramatsu:2022fgn} implementing the new parameter $\alpha_{V}$.

Here is the organisation of the paper. In section~\ref{sec:introEFTofDE}, we introduce the unified EFT of DE which accommodates both the scalar-tensor and vector-tensor theories. We also discuss the dictionary between the so-called alpha parameters~$\alpha_i$ and the Horndeski/generalised Proca Lagrangians. In section~\ref{sec:evolution_equation}, we describe both the background and the perturbation equations of motion, especially focusing on how the new parameter~$\alpha_V$ and the consistency relations affect these equations.
Subsequently, we demonstrate the impact of the new parameter~$\alpha_{V}$ and the consistency relations in the CMB power spectrum in section~\ref{sec:demonstrations}. 
Then, in section \ref{sec:conclusion}, we present our conclusions. In the \hyperref[sec:matter]{Appendix}, we briefly review the basics of our EFT formulation.

\section{Unified EFT of dark energy}\label{sec:introEFTofDE}

\subsection{EFT action}
\label{sec:EFTaction}

Despite the name of the EFT of {\it dark energy} (EFT of DE), the commonly-used EFT of DE can only deal with the DE scenario driven by a scalar field~\cite{Gubitosi:2012hu}. The EFT approach for vector-tensor theories was recently formulated in~\cite{Aoki:2021wew}, and it serves as a unified framework of DE scenarios that accommodates not only vector-tensor theories but also scalar-tensor theories. Incorporating vector-tensor theories into the EFT of DE requires two additional ingredients, a new phenomenological function~$\alpha_V(t)$ and consistency relations, which we shall explain in this section. The details of the EFT formulation are found in~\cite{Aoki:2021wew} (see also the \hyperref[sec:matter]{Appendix} for a brief review). 

For simplicity, we focus on linear perturbations around the spatially flat Friedmann-Lema\^itre-Robertson-Walker (FLRW) background without vector-type perturbations, although the formulation of the vector-tensor EFT is not limited to this situation. When ignoring the vector perturbations, we can simply adopt the ADM form of the metric
%
\begin{align}
\D s^2=-N^2\D t^2+h_{ij}(\D x^i + N^i \D t)(\D x^j + N^j \D t)
\,,
\label{eq:ADM_metric}
\end{align}
%
as the basis of the EFT formulation even in the vector-tensor theories. The extrinsic and spatial curvatures are denoted by $K_{ij}$ and $\spatialR_{ijkl}$, respectively. As for the background, we use the spatially flat FLRW spacetime,
%
\begin{align}
N=\bar{N}(t)\,, \qquad N^i=0\,, \qquad h_{ij}=a^2(t)\delta_{ij}\,,
\label{FLRW}
\end{align}
%
and the Hubble expansion rate is denoted by $H\equiv \dot{a}/a$, with $\dot{a}\equiv {\rm d}a/(\bar{N}{\rm d}t)$. In the following, we do not specify the background time coordinate so that one can easily work with the convenient time coordinate such as the cosmic time ($\bar{N}=1$) or the conformal time ($\bar{N}=a$). We then introduce the perturbations of the lapse function and the extrinsic curvature as
%
\begin{align}
\delta N \equiv N- \bar{N}\,, \qquad
\delta K^{i}{}_{j}\equiv K^{i}{}_{j} - H(t) \delta^{i}{}_{j}\,.
\label{def_perturbations}
\end{align}
%
Then, the quadratic action for the scalar and tensor perturbations in the momentum space is computed by the following effective action in the unitary gauge~\cite{Aoki:2021wew} (see the \hyperref[sec:matter]{Appendix} for the details):
%
\begin{align}
    S_2 = \int &\frac{\D t \D^3k}{(2\pi)^3} \bar{N}a^3  \frac{M^2}{2}
    \nn
    \times \Biggl[ & 
    (1+\tilde{\alpha}_H) \frac{\delta_1 N}{\bar{N}} \delta_1 \! \spatialR + 4 H \tilde{\alpha}_B \frac{\delta_1 N}{\bar{N}} \delta_1 K
    +\delta_1 K^{i}{}_{j} \delta_1 K^{j}{}_{i}
    -(1+\tilde{\alpha}_B^{\rm GLPV}) (\delta_1 K)^2 
    \nn
    &
    + \tilde{\alpha}_K H^2 \left(\frac{\delta_1 N}{\bar{N}}\right)^2
    +(1+\alpha_T) \delta_2\bigg( \spatialR \frac{\sqrt{h}}{a^3}\, \bigg)  + \frac{\tilde{\alpha}_M^{\rm GC}}{H^2} (\delta_1 \! \spatialR)^2
    + \frac{\tilde{\alpha}_B^{\rm GC} }{H} \delta_1 K  \delta_1 \! \spatialR + \cdots \Biggl] 
    + S^{\rm m}_2 \,,
    \label{LEFT_alpha}
\end{align}
%
where the ellipsis stands for higher-derivative terms that we ignore in the present paper. Here, $\delta_1 \mathcal{Q}$ and $\delta_2 \mathcal{Q}$ are the linear- and quadratic-order parts of the perturbations~$\delta \mathcal{Q}$. 
The action~$S^{\rm m}_2$ is the quadratic action for the matter fields which is given by
%
\begin{align}
S^{\rm m}_2 = \delta_2 S^{\rm m} + \delta_2\int \D^4 x \bar{N}\sqrt{h}\left( \bar{\rho}_{\rm m} \frac{\delta N}{\bar{N}} - \bar{p}_{\rm m} \right)
\,,
\label{eq:quad_matter}
\end{align}
%
where  $S^{\rm m}$ is the full matter action. The second term arises from the ``background part'' of the effective action (see the \hyperref[sec:matter]{Appendix}). 
Several comments are in order. 
\begin{itemize}
\item {\bf EFT inputs.} The coefficients~$\{ M^2, \tilde{\alpha}_H, \tilde{\alpha}_B, \tilde{\alpha}_B^{\rm GLPV}, \tilde{\alpha}_K, \alpha_T, \tilde{\alpha}^{\rm GC}_M, \tilde{\alpha}^{\rm GC}_B\}$ parameterise the quadratic action for the linear perturbations. Those with a tilde are functions of the time~$t$ and the momentum~$k$, and the momentum dependence is fixed by
%
\begin{align}
    \begin{split}
    \tilde{\alpha}_B(t,k)& = [1-\mathcal{G}(t,k)]\alpha_B(t)\,, \\
    \tilde{\alpha}_K(t,k) &= [1-\mathcal{G}(t,k)]\alpha_K(t) \,, \\
    \tilde{\alpha}_H(t,k) &= [1-\mathcal{G}(t,k)]\alpha_H(t) + \mathcal{G}(t,k) \alpha_T(t)
    \,, \\
    \tilde{\alpha}_B^{\rm GLPV}(t,k)  &= \alpha_B^{\rm GLPV}(t) + 4 \mathcal{G}(t,k) \frac{\alpha_B^2(t)}{\alpha_K(t)}
    \,, \\
    \tilde{\alpha}_M^{\rm GC}(t,k) &= \alpha_M^{\rm GC}(t)-\frac{1}{4}\mathcal{G}(t,k) \frac{[\alpha_H(t)-\alpha_T(t)]^2}{\alpha_K(t)}
    \,, \\
    \tilde{\alpha}_B^{\rm GC}(t,k) &= \alpha_B^{\rm GC}(t)-2 \mathcal{G}(t,k) \frac{\alpha_B(t)[\alpha_H(t)-\alpha_T(t)]}{\alpha_K(t)}
    \,,
    \end{split}
    \label{alpha_relation}
\end{align}
%
with
%
\begin{align}
    \mathcal{G}(t,k)\equiv \frac{\alpha_V \alpha_K}{\alpha_V \alpha_K + k^2/(a^2H^2)}
    \,,\label{eq:def_G}
\end{align}
%
where $\alpha_K \neq 0$ has been assumed.
Therefore, on top of the background dynamics, the essential inputs of the EFT are the $5+3+1$ time-dependent functions,
%
\begin{align}
    M^2(t)\,, \quad
    \alpha_K(t)\,, \quad 
    \alpha_B(t)\,, \quad
    \alpha_T(t)\,, \quad
    \alpha_H(t)\,, 
    \label{Malpha} \\
    \alpha_B^{\rm GLPV}(t)\,, \quad \alpha_M^{\rm GC}(t)\,, \quad \alpha_B^{\rm GC}(t)\,, 
    \label{alphaGC}\\
    \alpha_V(t) 
    \label{alphaV}
    \,.
\end{align}
%
The first five functions~\eqref{Malpha} appear in the standard EFT of DE: $M$ is the effective Planck mass for gravitational waves (GWs), $\alpha_K$ is kineticity, $\alpha_B$ is kinetic gravity braiding, $\alpha_T$ describes the deviation of the GW speed from the speed of light, and $\alpha_H$ characterises the deviation from the Horndeski theory (generalised Proca theory in the vector-tensor case)~\cite{Bellini:2014fua}.\footnote{More precisely, the function~$\alpha_H$ parameterises the effect of the GLPV theory. The extension to degenerate higher-order scalar-tensor theories requires additional functions~$\{\beta_1,\beta_2,\beta_3\}$ (or more) with certain constraints associated with the degeneracy conditions~\cite{Langlois:2017mxy,Takahashi:2023jro}.} It is convenient to introduce the parameter representing the evolution rate of the effective Planck mass,
%
\begin{align}
    \alpha_M(t) \equiv \frac{1}{H} \frac{\D \ln M^2(t)}{\bar{N} \D t}\,.
    \label{def_alphaM}
\end{align}
%
The function~$\alpha_B^{\rm GLPV}$ describes the departure from the GLPV theory. 
The functions~$\alpha_M^{\rm GC}$ and $\alpha_B^{\rm GC}$ (as well as $\alpha_B^{\rm GLPV}$) appear in the scenario of (gauged) ghost condensate. The non-vanishing functions~\eqref{alphaGC} lead to a $k^4$ term in the dispersion relation of the scalar mode and are responsible for ghost condensate and the scordatura theory~\cite{Motohashi:2019ymr,Gorji:2020bfl,Gorji:2021isn,DeFelice:2022xvq}. Finally, the function~$\alpha_V$ characterises the vectorial nature and determines the non-local $k$-dependence of the effective action~\eqref{LEFT_alpha} through the relations~\eqref{alpha_relation}. The scalar-tensor theory is recovered by setting $\alpha_V=0$ and the vector-tensor theory corresponds to $\alpha_V>0$, where the negative region~$\alpha_V<0$ is excluded by the ghost-free condition of the vector-type perturbations~\cite{Aoki:2021wew}.

\item {\bf Stability conditions.} We require that the perturbations are stable at sub-horizon scales; that is, there is neither ghost nor gradient instability. The stability conditions for tensor perturbations and scalar perturbations are respectively given by~\cite{Aoki:2021wew}
%
\begin{align}
M^2&>0 \,, \quad
1+\alpha_T >0 \,, \label{stability_tensor} \\
\alpha_K+6\alpha_B^2&>0 \,, \quad
V_S>0
\,, \label{stability_scalar}
\end{align}
%
where
%
\begin{align}
 V_S &\equiv  4\alpha_V\mathcal{A}^2
    +2(1+\alpha_B)\mathcal{A}
    \nn
    &\quad -2(1+\alpha_H)^2\left[ \frac{1+\alpha_B}{1+\alpha_H}\left(\frac{\dot{H}}{H^2}-\alpha_M \right)+\frac{1}{H}\frac{\D}{\bar{N}\D t} \left(\frac{1+\alpha_B}{1+\alpha_H} \right) + \sum_a \frac{\bar{\rho}_a+\bar{p}_a}{2M^2H^2} \right]
    \,, \label{def_VS}
\end{align}
%
with
%
\begin{align}
    \mathcal{A}\equiv \alpha_H-\alpha_T-\alpha_B(1+\alpha_T)\,. \label{def_calA}
\end{align}
%
Here, $\bar{\rho}_a$ and $\bar{p}_a$ are the background energy density and the pressure of the matter species~$a$, and the summation is over all matter species. 
As we have already mentioned, the ghost-free condition for vector perturbations in the vector-tensor theories requires
%
\begin{align}
\alpha_V > 0
\,. \label{stability_vector}
\end{align}
%
The condition for no gradient instability gives an additional condition on the vector-tensor EFT~\cite{Aoki:2021wew}. However, this condition is not relevant for the scalar and tensor sectors and therefore is not presented here.

\item {\bf Size of $\alpha_i$.} Although the $\alpha$-parameters are all dimensionless, there may be a hierarchy among them. The functions~$\{\alpha_K, \alpha_B, \alpha_T, \alpha_H, \alpha_M  \}$ are normalised with respect to the size of modifications of gravity on cosmological scales. If the modification of gravity is of order unity, we would have
\begin{align}
\alpha_K, \alpha_B, \alpha_T, \alpha_H, \alpha_M = \mathcal{O}(1)
\,.
\end{align}
Conventionally, the functions~$\{\alpha^{\rm GLPV}_B, \alpha^{\rm GC}_M, \alpha^{\rm GC}_B \}$ are similarly normalised but they appear in the coefficients of the higher-derivative operators. Therefore, their values must be suppressed by the cutoff scale~$\cutoff$, reading 
\begin{align}
\alpha^{\rm GLPV}_B, \alpha^{\rm GC}_M, \alpha^{\rm GC}_B = \mathcal{O}(H^2/\cutoff^2) \ll 1
\,.
\label{alphaGC_values}
\end{align}
The function~$\alpha_V$ is defined in \eqref{alpha-V} as the ratio of the (dimensionful) gauge coupling of the vector field to the gravitational coupling and determines the scale at which the behaviour of $\mathcal{G}(t,k)$ changes qualitatively~\cite{Aoki:2021wew}. The scale appears at the horizon scale for $\alpha_V=\mathcal{O}(1)$ and at a sub-horizon scale for $\alpha_V \gg 1$. On the other hand, the vectorial effects can be ignored when $\alpha_V \ll 1$, which is regarded as the scalar-tensor limit.

\item {\bf Consistency relations.} In the generic scalar-tensor theories $(\alpha_V=0)$, the coefficients \eqref{Malpha} and \eqref{alphaGC} can have independent time-dependence. On the other hand, the shift-symmetric scalar-tensor theories $(\alpha_V=0)$ and the vector-tensor theories $(\alpha_V>0)$ have to satisfy the following relations to be consistent with the requirements of the global/local shift symmetry:
\begin{align}
    2M^2 \dot{H} \frac{\alpha_K + 6\alpha_B^2}{\alpha_K} + \frac{\mathcal{J}(t)}{a^3} \frac{\alpha_K-6\alpha_B}{\alpha_K} + \sum_a \left(\bar{\rho}_a+\bar{p}_a \right)&\simeq 0
    \,, \label{consistency1_alpha} \\
    H\alpha_M(1+\alpha_T)+\dot{\alpha}_T + \frac{6}{H} \frac{ \alpha_H-\alpha_T}{\alpha_K} \left(\dot{H} \alpha_B - \frac{1}{2M^2} \frac{\mathcal{J}(t)}{a^3}\right)&\simeq 0
    \,. \label{consistency2_alpha}
\end{align}
Note that the consistency relations~\eqref{consistency1_alpha} and \eqref{consistency2_alpha} are approximate equalities because higher-derivative corrections are ignored under the assumptions~\eqref{alphaGC_values}. In the consistency relations, we have introduced a new time-dependent function~$\mathcal{J}(t)$ satisfying
\begin{align}
\mathcal{J} (t) =
\begin{cases}
\dot{\phi}(t) J_0 & \text{for shift-symmetric scalar-tensor theories}~(\alpha_V=0)
\\
0 & \text{for vector-tensor theories}~(\alpha_V>0)
\end{cases}
\,, \label{calJ}
\end{align}
where $J_0$ is the charge associated with the shift-symmetry. The function~$\phi(t)$ corresponds to the background scalar field of the shift-symmetric scalar-tensor theories. When we choose the time coordinate such that $\phi=t$, the function~$\mathcal{J}(t)$, is given by $\mathcal{J}(t)=J_0/\bar{N}(t)$, which recovers the expressions for the consistency relations obtained in~\cite{Aoki:2021wew} on identifying $J_0$ with $2c_0$ in \cite{Aoki:2021wew}. One would generically expect that $\mathcal{J}/a^3$ is a decreasing function in time in an expanding universe. The function $\mathcal{J} $ describes how the background approaches the attractor phase $\mathcal{J}/a^3 \to 0$. 
\end{itemize}

In summary, the effective action~\eqref{LEFT_alpha} describes the scalar and tensor perturbations in not only DE scenarios driven by a scalar field but also those driven by a vector field.
There are three different classes of theories, i.e., generic scalar-tensor, shift-symmetric scalar-tensor, and vector-tensor theories, and they are distinguished by the phenomenological function~$\alpha_V(t)$ and the consistency relations~\eqref{consistency1_alpha} and \eqref{consistency2_alpha}. We summarise the classification in Table~\ref{table_EFTs}.

\begin{table}[t]
  \caption{Classification of the EFTs.}
  \label{table_EFTs}
  \centering
    \begin{tabular}{lcc}
  \hline 
   Theories & $\alpha_V$ & Consistency relations~\eqref{consistency1_alpha} \& \eqref{consistency2_alpha} \\
   \hline \hline
   Genetic scalar-tensor theories & $~\alpha_V=0~$ & Not required \\
   Shift-symmetric scalar-tensor theories & $\alpha_V=0$ & Required \\
   Vector-tensor theories & $\alpha_V >0 $ & Required with $\mathcal{J} =0$ \\
   \hline
  \end{tabular}
\end{table}


\subsection{Dictionary within Horndeski/generalised Proca family}
\label{sec:dictionary}

As a reference, we give background equations of motion in a subclass of Horndeski/generalised Proca theories. In such classes of theories, the $\alpha$-parameters satisfy
\begin{align}
\alpha_H= \alpha_B^{\rm GLPV} = \alpha_M^{\rm GC} = \alpha_B^{\rm GC} = 0
\,,
\label{Hclass}
\end{align}
and the explicit relations between the other EFT coefficients and the Lagrangian are summarised below.
The derivations of these results are found in~\cite{Aoki:2021wew} for the generalised Proca theory and in~\cite{Bellini:2014fua, Gleyzes:2014rba, Gleyzes:2014qga} for the Horndeski theory, respectively.\footnote{We follow the convention of \cite{Gleyzes:2014rba,Gleyzes:2014qga} in which the definition of $\alpha_B$ is slightly different from that in~\cite{Bellini:2014fua}.}

\subsubsection{Horndeski theory}
The action of the Horndeski theory up to quadratic in second derivatives of the scalar field $\phi$ is given by~\cite{Horndeski:1974wa, Deffayet:2011gz, Kobayashi:2011nu}
\begin{align}
S_H=\int \D^4 x\sqrt{-g}
\{ G_2(\phi,X)+G_3(\phi,X)\Box \phi + G_4(\phi,X)R+G_{4X}[(\Box \phi)^2-\nabla_{\mu}\nabla_{\nu}\phi \nabla^{\mu}\nabla^{\nu} \phi ] \} \,,
\label{LagH}
\end{align}
where $G_2, G_3, G_4$ are arbitrary functions of $\phi$ and $X\equiv -\nabla_{\mu}\phi \nabla^{\mu} \phi/2$. We shall use the notation that the subscripts denote derivatives with respect to the specified variables, e.g., $G_{2\phi}=\partial G_2/\partial \phi, G_{2\phi\phi}=\partial^2 G_2/\partial \phi^2, G_{4X}=\partial G_4/\partial X$.
When the shift symmetry is imposed, the functions must depend on $X$ only and then the functions with the subscript~$\phi$ vanish.

Let us consider the flat FLRW background~\eqref{FLRW} with the homogeneous background configuration of the scalar field~$\phi=\phi(t)$. The background equations of motion are given by
\begin{align}
G_2-2XG_{2X}-2XG_{3\phi}+6(2X)^{1/2} H (XG_{3X}+2XG_{4\phi X}+ G_{4\phi}) 
+6H^2(G_4-4X G_{4X}-4X^2 G_{4XX}) &= \sum_a \bar{\rho}_a
\,,
\label{FriH1}
\\
G_2+2X G_{3\phi}+(6H^2+4\dot{H})(G_4-2X G_{4X}) + 4XG_{4\phi\phi} - 4H\dot{X}(G_{4X}+2XG_{4XX})&
\nn
+(2X)^{1/2}\left[ 4H (G_{4\phi}-2G_{4\phi X})+ \frac{\dot{X}}{X} ( X G_{3X}+G_{4\phi} +2 X G_{4\phi X}) \right]&=-\sum_a \bar{p}_a
\,,
\label{FriH2}
\end{align}
and
\begin{align}
\dot{J}+3HJ = G_{2\phi}+2XG_{3\phi\phi}+(2X)^{1/2} \dot{X} G_{3\phi X} + 6(2H^2 +\dot{H}) G_{4\phi} - 6H(\dot{X}+2HX)G_{4\phi X}
\,,
\label{scalar_eom}
\end{align}
with
\begin{align}
J\equiv (2X)^{1/2}\left[ G_{2X}  + 2G_{3\phi}+6H^2(G_{4X} +2X G_{4XX}) \right]- 6HXG_{3X}-12 HX G_{4\phi X}
\,.
\end{align}
Here, $X$ and the functions~$G_i$ are evaluated on the background, i.e.,
\begin{align}
X(t)=\frac{1}{2} \dot{\phi}^2(t), \qquad 
G_i=G_i(\phi(t), X(t))
\,.
\end{align}
As for the linear perturbations, the $\alpha$-parameters for the Horndeski theory~\eqref{LagH} are given by
\begin{align}
M^2&= 2(G_4-2XG_{4X})
\,, \label{M2_H} \\
H^2M^2 \alpha_K &=- 12 (2X)^{1/2} X H\left[  G_{3X}+X G_{3XX}+3 G_{4\phi X}+ 2X G_{4\phi XX}\right]
\nn
&\quad +2X\left[ G_{2X}+2X G_{2XX}+2G_{3\phi}+2X G_{3\phi X}+6H^2(G_{4X}+8X G_{4XX}+4X^2 G_{4XXX}) \right]
\,, \\
HM^2\alpha_B &= (2X)^{1/2}(X G_{3X}+G_{4\phi}+2X G_{4\phi X})-4H X(G_{4X}+2X G_{4XX})
\,, \\
M^2 \alpha_T &=4XG_{4X}
\,. \label{alphaT_H}
\end{align}

In the shift-symmetric theories, the field equation~\eqref{scalar_eom} can be integrated once to give $J=J_0/a^3$ with the integration constant~$J_0$, which is nothing but the charge associated with the shift-symmetry in \eqref{calJ}. One can confirm that the consistency relations~\eqref{consistency1_alpha} and \eqref{consistency2_alpha} indeed hold under the background equations~\eqref{FriH1}--\eqref{scalar_eom} in the shift-symmetric theories.

\subsubsection{Generalised Proca theory}
The action of a subclass of the generalised Proca theory with a vector field $A_{\mu}$ is given by~\cite{Heisenberg:2014rta}
\begin{align}
S_{\rm GP}=\int \D^4x\sqrt{-g} \left\{G_2(X,F,Y)+G_3(X)\nabla_\mu A^\mu+G_4(X)R+G_{4X}[ (\nabla_\mu A^\mu)^2-\nabla_\mu A_\nu \nabla^\nu A^\mu] \right\}\,, \label{GP}
\end{align}
with
\begin{align}
X\equiv -\frac{1}{2}A_\mu A^\mu\,, \qquad
    F\equiv -\frac{1}{4}F_{\mu\nu}F^{\mu\nu}\,, \qquad
    Y\equiv A^\mu A^\nu F_{\mu\alpha}F_{\nu}{}^{\alpha}\,.
\end{align}
We use the same notation as in the case of Horndeski theory~\eqref{LagH}.

The cosmological solution is found under the ansatz~$A_{\mu}=(\bar{A}_0(t),0,0,0)$ with \eqref{FLRW}. The background equations are
\begin{align}
G_2-2XG_{2X}+3(2X)^{3/2} H G_{3X} 
+6H^2(G_4-4X G_{4X}-4X^2 G_{4XX}) &= \sum_a \bar{\rho}_a
\,,
\label{FriGP1}
\\
G_2+(6H^2+4\dot{H})(G_4-2X G_{4X}) - 4H\dot{X}(G_{4X}+2XG_{4XX})
+(2X)^{1/2}\dot{X} G_{3X} &=-\sum_a \bar{p}_a
\,,
\label{FriGP2}
\\
J\equiv (2X)^{1/2}\left[ G_{2X}  +6H^2(G_{4X} +2X G_{4XX}) \right]- 6HXG_{3X}&=0
\,.
\label{vector_eom}
\end{align}
Since $F_{\mu\nu}$ identically vanishes under the ansatz~$A_{\mu}=(\bar{A}_0(t),0,0,0)$, not only $G_3, G_4$ but also the function~$G_2$ are regarded as the single-variable functions of $X(t)$ when they are evaluated on the background:
\begin{align}\label{X-GP}
X(t)=\frac{1}{2}\frac{\bar{A}_0^2(t)}{\bar{N}^2(t)}\,, \qquad G_i=G_i(X(t))
\,.
\end{align}
The $\alpha$-parameters for the generalised Proca theory are computed by
\begin{align}
M^2&= 2(G_4-2XG_{4X}) \label{M2_GP}
\,, \\
H^2M^2 \alpha_K &=- 12 (2X)^{1/2} X H\left[  G_{3X}+X G_{3XX} \right]
\nn
&\quad +2X\left[ G_{2X}+2X G_{2XX}+6H^2(G_{4X}+8X G_{4XX}+4X^2 G_{4XXX}) \right]
\,, \\
HM^2\alpha_B &= \frac{1}{2}(2X)^{3/2} G_{3X}-4H X(G_{4X}+2X G_{4XX})
\,, \\
M^2 \alpha_T &=4XG_{4X}
\,, 
\end{align}
and
%
\begin{align}
\alpha_V=\frac{M^2}{2X G_{2F}+8X^2 G_{2Y}}
\,. \label{alphaV_GP}
\end{align}
%
The background equations and the formulae for $\{M^2, \alpha_K, \alpha_B, \alpha_T \}$ in the generalised Proca theory agree with those in the Horndeski theory with the shift symmetry by setting $J_0=0$.\footnote{One may obtain a slightly different expression for $\alpha_K$ using the dictionary given in~\cite{Aoki:2021wew}. Here, we have added the background equation~$J=0$ so that the expression for $\alpha_K$ agrees with that in the Horndeski theory: $\alpha_K= \alpha_K^{\text{\cite{Aoki:2021wew}}}+(2X)^{1/2}J H^2 M^2$.}


\section{Evolution equations}\label{sec:evolution_equation}

\subsection{Background}

In the concrete model approach, the background dynamics of the Universe and the DE field are obtained by solving the background equations~\eqref{FriH1}--\eqref{scalar_eom} or \eqref{FriGP1}--\eqref{vector_eom} combined with the equations of motion for the matter fields. Note that all the equations are not independent by virtue of the Bianchi identity. The evolutions of perturbations are then solved with the $\alpha$-parameters computed according to \eqref{M2_H}--\eqref{alphaT_H} or \eqref{M2_GP}--\eqref{alphaV_GP}.

In the pure EFT approach, on the other hand, all the time-dependent functions, including the dynamics of the Hubble expansion rate~$H(t)$, are input parameters to realise a designed cosmological history. Although one can directly fix the time-dependence of $H(t)$ in principle, it would be more convenient to parameterise the dynamics of $H$ by use of the notion of the effective DE component by following the strategy of {\tt EFTCAMB}~\cite{Hu:2013twa, Raveri:2014cka}. 

Let us introduce the effective energy density and pressure of DE via
%
\begin{align}
3M_{\rm cos}^2H^2&=\bar{\rho}_{\rm DE} + \sum_a \bar{\rho}_a
\,, \label{def_Fri1} \\
2M_{\rm cos}^2\dot{H}&=-\left[(\bar{\rho}_{\rm DE}+ \bar{p}_{\rm DE}) + \sum_a( \bar{\rho}_a+ \bar{p}_a)\right]\,,
\label{def_Fri2}
\end{align}
%
where $\bar{\rho}_a$ and $\bar{p}_a$ are the energy density and the pressure for the ordinary matter components, respectively. 
Here, $M^2_{\rm cos}(t)$ is the Planck mass for the background dynamics which can be time-dependent. In general, there are no unique definitions of the cosmological Planck mass and the effective DE component in modified gravity theories because there is no clear separation between the ``gravity'' part and the ``DE'' part in the background equations [see Eqs.~\eqref{FriH1}, \eqref{FriH2}, \eqref{FriGP1}, \eqref{FriGP2}]. For instance, one can use $M^2_{\rm cos}=M_{\rm Pl}^2$ with the measured Planck mass~$M_{\rm Pl}^2$. However, we do not specify the cosmological Planck mass~$M^2_{\rm cos}$ for a while.
By using the conservation laws of the ordinary matter components, Eqs.~\eqref{def_Fri1} and \eqref{def_Fri2} yield
%
\begin{align}
\dot{\bar{\rho}}_{\rm DE}+3H(\bar{\rho}_{\rm DE}+ \bar{p}_{\rm DE}) = 3H^2 \frac{\D}{\bar{N}\D t} M_{\rm cos}^2\,.
\label{DE_conserve}
\end{align}
%
The density parameters and the equation-of-state parameters are defined by
%
\begin{align}
\Omega_A \equiv \frac{\rho_A}{3M^2_{\rm cos}H^2}\,, \qquad w_A\equiv \frac{p_A}{\rho_A}
\,,
\end{align}
%
with $A=\{a,{\rm DE} \}$. Here, the definitions of $\Omega_A$ and $w_{\rm DE}$ depend on the choice of $M_{\rm cos}^2$.

Instead of fixing $H$, we can use $w_{\rm DE}$ and $M^2_{\rm cos}$ as input parameters of the background dynamics in the pure EFT approach. In this case, the equations~\eqref{def_Fri1} and \eqref{DE_conserve} are regarded as the evolution equations for $H$ and $\rho_{\rm DE}$. 
Note that there are two inputs $\{ w_{\rm DE}, M^2_{\rm cos} \}$ for the background dynamics while the original input of the pure EFT approach at the background level is $H$ only. The appearance of the additional input parameter is due to the ambiguity of the definition of $M_{\rm cos}^2$. There are at least two convenient choices to reduce the redundant input parameter:
\begin{enumerate}
\item $M^2_{\rm cos}=M_{\rm Pl}^2$. This would be one of the simplest parameterisations of the background. The ``continuity equation'' of the DE~\eqref{DE_conserve} takes a simple form and can be integrated as a function of the scale factor:
%
\begin{align}
\rho_{\rm DE}(a)=\rho_{{\rm DE},0}\exp\left[ -3 \int^{\ln a}_{\ln a_0}[1+w_{\rm DE}(a')] \D \ln a' \right] \,,
\end{align}
%
where $\rho_{{\rm DE},0}$ and $a_0$ are the DE density and the scale factor at the present time, respectively. In this case, all the modifications of the Friedman equation, including the time-dependent change of the gravitational interaction, are interpreted as the effective DE component.

\item $M^2_{\rm cos}=M^2(t)$. The cosmological Planck mass is identified with the effective Planck mass appearing in the perturbation equations in the $\alpha$-basis. Recall that $M^2(t)$ is the Planck mass for the GWs. This parameterisation is particularly useful for dealing with the consistency relation~\eqref{consistency1_alpha} as discussed in Sec.~\ref{sec:consistency}.
\end{enumerate}

\subsection{Perturbations}

The EFT action~\eqref{LEFT_alpha} is defined in the unitary gauge in which the scalar mode of DE is eaten by the spacetime metric. On the other hand, it is possible to introduce the St{\"u}ckelberg field to restore the general covariance and then move to another convenient gauge. We define the metric functions in \eqref{eq:ADM_metric} as
%
\begin{align}
 N=1+\delta N\,, \qquad
 N_{i} = a\partial_i\beta\,,
 \qquad h_{ij} = a^2(1+2\zeta)\delta_{ij}+a^2\left(\partial_i\partial_j-\frac{1}{3}\delta_{ij}\partial^k \partial_k\right)E\,.
\end{align}
%
Shifting the time coordinate, $t \to t+T(t,\xx)$, to move to the Newtonian gauge, we obtain
%
\begin{align}
 \Psi = \delta N-\dot{T}\,, \qquad
 0=\beta + \frac{1}{a}T\,, \qquad
 \Phi = -\zeta+HT\,,
\end{align}
%
where $E$ remains unchanged.
Later, we will set $E=0$ by using the freedom of the spatial coordinates but for now keep it as a placeholder.
The second equation gives $T = -a\beta$, and we then find
%
\begin{align} 
 \Psi = \delta N + a(\dot{\beta}+H\beta)\,, \qquad
 \Phi = -\zeta - aH\beta\,.
\end{align}
As a result, we have moved to the Newtonian gauge:
%
\begin{align}
\D s^2 = -(1+2\Psi)\D t^2 + a^2 \left[(1 - 2\Phi)\delta_{ij} + \left(\partial_i\partial_j-\frac{1}{3}\delta_{ij}\partial^k \partial_k\right)E\right]\D x^i\D x^j\,.
\end{align}
As a scalar degree of freedom manifests in this gauge, we define $\Pi \equiv -HT = aH\beta$.

One can expand the action~\eqref{LEFT_alpha} in terms of the Newtonian-gauge perturbation variables to find the quadratic Lagrangian, which we denote by $\mathcal{L}_2$.
The equations of motion for the perturbations can then be obtained by varying the quadratic Lagrangian~$\mathcal{L}_2$ with respect to the metric perturbations and the scalar perturbation, formally given as
%
\begin{align}
\mathcal{E}_Q \equiv \frac{\partial\mathcal{L}_2}{\partial Q} - \partial_\mu\frac{\partial\mathcal{L}_2}{\partial (\partial_\mu Q)} + \partial_\mu\partial_\nu\frac{\partial\mathcal{L}_2}{\partial (\partial_\mu \partial_\nu Q)} + \cdots\,,
\end{align}
%
where $Q \in \{\Psi, \Phi, E, \Pi\}$.\footnote{Note that this manipulation (i.e., gauge fixing at the Lagrangian level) is legitimate as one can move to the Newtonian gauge by complete gauge fixing~\cite{Motohashi:2016prk}.}
As a result, we obtain
%
\begin{align}
 \mathcal{E}_{\Psi}           &= \mathcal{E}_{\Psi}(\Psi,\Phi,\dot{\Phi},\Pi,\dot{\Pi};\delta\rho)\,, \label{eq:EL1}\\                                         \mathcal{E}_{\Phi}           &= \mathcal{E}_{\Phi}(\Psi,\dot{\Psi},\Phi,\dot{\Phi},\ddot{\Phi},\Pi,\dot{\Pi},\ddot{\Pi};\delta p)\,, \label{eq:EL2}\\
 \mathcal{E}_{E}                 &= \mathcal{E}_{E}(\Psi,\Phi,\dot{\Phi},\Pi,\dot{\Pi};\delta\Sigma)\,, \label{eq:EL4}\\
 \mathcal{E}_{\Pi} &= \mathcal{E}_{\Pi}(\Psi,\dot{\Psi},\Phi,\dot{\Phi},\ddot{\Phi},\Pi,\dot{\Pi},\ddot{\Pi})\,, \label{eq:EL5}
\end{align}
%
where $\delta\rho$, $\delta p$, and $\delta\Sigma$ are the perturbations of energy density, pressure, and anisotropic stress, respectively. 
In the matter sector, we assume the cold dark matter, baryons, photons, and massless neutrinos.
A linear combination of them and $\dot{\mathcal{E}}_{E}$ yields a set of equations to be solved,
%
\begin{align}
 \ddot{\Pi} &= F_1(\Phi, \Pi, \dot{\Pi}; \delta_c, \delta_b, \Theta_{A0}, \Theta_{A1}, \Theta_{A2}, \Theta_{A3})\,, \label{eq:pert1} \\
 \dot{\Phi} &= F_2(\Phi, \Pi, \dot{\Pi}; \delta_c, \delta_b, \Theta_{A0}, \Theta_{A2})\,, \label{eq:pert2} \\
 \Psi &= F_3(\Phi, \Pi, \dot{\Pi}; \delta_c, \delta_b, \Theta_{A0}, \Theta_{A2})\,, \label{eq:pert3}
\end{align}
%
where $\delta_{c}$ and $\delta_{b}$ are the density fluctuations of the cold dark matter and baryons, respectively, $\Theta_{A\ell}$ is the temperature anisotropies of photons ($A=\gamma$) and neutrinos ($A=\nu$) with the angular multipole~$\ell$. We do not show their lengthy concrete expressions as they are unimportant in the later discussion. The form of the equations of motion is formally the same as that of the EFT of scalar-tensor theories. The only change is that the coefficients~$\alpha_{i}(t)$ get modified to $\tilde{\alpha}_i(t,k)$ by $\alpha_V(t)$ as presented in Eq.~(\ref{alpha_relation}). Note that our numerical code uses the conformal time (${\bar N}=a$) instead of the proper time (${\bar N}=1$).

\subsection{Effective gravitational coupling}
Before studying the CMB spectrum, we briefly comment on how the gravitational law changes in the EFT, especially the role of $\alpha_V$. For simplicity, we only consider the non-relativistic matter in this subsection.
In the classes of the Horndeski/generalised Proca theories with \eqref{Hclass}, the Poisson equation is easily derived under the quasi-static approximation in a sufficiently small scale~$k^2/a^2 \gg \alpha_K \alpha_V H^2$. The equation of motion for the St{\"u}ckelberg field~$\Pi$ is algebraically solved under the quasi-static approximation. We then obtain the following gravitational field equations~\cite{Aoki:2021wew}:
%
\begin{align}
    \frac{k^2}{a^2} \Psi &= -\mu (t)4\pi G \bar{\rho}_{\rm m} \Delta_{\rm m}
    \label{Poisson_eq} \,, \\
    \eta (t) & = \frac{\Phi}{\Psi}
    \label{slip_para}
    \,,
\end{align}
%
where $8\pi G=1/M^2_{\rm Pl}$ is the observed gravitational constant, and $\bar{\rho}_{\rm m}$ and $\Delta_{\rm m}$ are the background energy density and the comoving density contrast of the non-relativistic matter, respectively. The effective gravitational coupling~$\mu$ and the slip parameter~$\eta$ are given by
%
\begin{align}
    \mu(t) &=\frac{M^2_{\rm Pl}}{M^2}\left[ 1+ \alpha_T -2\alpha_T^2 \alpha_V + \frac{2}{V_S}( \alpha_M + \mathcal{A}-2\mathcal{A}\alpha_T \alpha_V)^2 \right]
    \,, \label{mu:H=0} \\
    \eta(t) &=\frac{1}{\mu} \frac{M^2_{\rm Pl}}{M^2}\left[ 1+ \frac{2(\mathcal{A}+\alpha_T)}{V_S (1+\alpha_T)}
    ( \alpha_M + \mathcal{A}-2\mathcal{A}\alpha_T \alpha_V) \right]
    \,, \label{eta:H=0}
\end{align}
%
where $\mathcal{A}$ is defined in Eq.~\eqref{def_calA}. Notice that, for Horndeski/generalised Proca theory, Eq.~(\ref{Hclass}) is satisfied.

It is now clear how the vectorial nature of DE $(\alpha_V>0)$ affects the modification of gravity. 
The stability conditions~\eqref{stability_tensor} and \eqref{stability_scalar} require that $1+\alpha_T$ and the last term inside the square brackets in \eqref{mu:H=0} are positive while the term~$-2\alpha_T^2 \alpha_V$ is negative. Therefore, the vector-tensor theories may weaken the gravitational force when $\alpha_T \neq 0$ as pointed out by~\cite{DeFelice:2016uil}.
In the case of $\alpha_T=0$ (corresponding to the luminal propagation of GWs), the negative term~$-2\alpha_T^2 \alpha_V$ is absent. In this case, Eqs.~\eqref{mu:H=0} and \eqref{eta:H=0} are simplified to be
\begin{align}
 \mu|_{\alpha_T=0} &=\frac{M^2_{\rm Pl}}{M^2}\left[ 1 + \frac{2(\alpha_B-\alpha_M )^2}{V_S} \right]
\,, \\
\eta|_{\alpha_T=0} &=\frac{1}{\mu} \frac{M^2_{\rm Pl}}{M^2}\left[ 1 + \frac{2\alpha_B(\alpha_B-\alpha_M)}{V_S} \right] 
\,,
\end{align}
with
\begin{align}
V_S=4\alpha_V \alpha_B^2 + (\text{$\alpha_V$-independent terms}) >0 \,.
\end{align}
Since we should have $V_S|_{\alpha_V>0} > V_S|_{\alpha_V=0}$, the non-zero $\alpha_V$ suppresses the enhancement of the effective gravitational constant due to the fifth force.  In particular, the Poisson equations get back to the GR forms in the limit $\alpha_V \to \infty$ even in the presence of the kinetic gravity braiding~$\alpha_B \neq 0$, where $M=M_{\rm Pl}$ is obtained as a consequence of the consistency relations as we will see shortly. 

While we have focused on the small scales in this subsection, we will also see a similar suppression effect of $\alpha_V$ in the CMB scales which will be studied in Sec.~\ref{sec:demonstrations}.

\subsection{Consistency relations}
\label{sec:consistency}

Finally, we elaborate on how the consistency relations~\eqref{consistency1_alpha} and \eqref{consistency2_alpha} impose non-trivial restrictions on the evolution of the system. The consistency relations arise as a requirement of the shift symmetry (either global or local). In general, the consistency relations involve the additional function~$\mathcal{J}$ associated with the global shift symmetry. However, the contribution of $\mathcal{J}$ is diluted due to the contribution~$a^{-3}$ as long as the growth of $\mathcal{J}$ is slower than $a^3$. We thus focus on the attractor phase in which $\mathcal{J} \to 0$. Note that the vector-tensor theories require $\mathcal{J}=0$. Therefore, by focusing on the attractor phase, the consistency relations are equally implemented in both the shift-symmetric scalar-tensor theories and the vector-tensor theories.
Assuming $\mathcal{J}=0$, the consistency relations are given by 
%
\begin{align}
    2M^2 \dot{H} \frac{\alpha_K + 6\alpha_B^2}{\alpha_K} + \sum_a \left(\bar{\rho}_a+\bar{p}_a \right)&= 0
    \,, \label{consistency1_alphaA} \\
    (1+\alpha_T)\frac{\D \ln M^2(t)}{\D t} +\dot{\alpha}_T + \frac{6\dot{H}}{H} \frac{ \alpha_B(\alpha_H-\alpha_T)}{\alpha_K} &= 0
    \,, \label{consistency2_alphaA}
\end{align}
%
where the approximate equalities have been replaced with the equalities for the code implementation. 

The ordinary matter components should satisfy the null energy condition~$\bar{\rho}_a+\bar{p}_a>0$. Let us assume $\dot{H} < 0$ at an early stage of the Universe so that the standard cosmic history (the matter/radiation dominant epoch) can be recovered.
Then, combining \eqref{consistency1_alphaA} with the stability conditions~$M^2>0$ and $\alpha_K+6\alpha_B^2>0$ (see Sec.~\ref{sec:EFTaction}), we obtain $\alpha_K>0$ at that time. Also, since we have assumed $\alpha_K \neq 0$, the parameter~$\alpha_K$ remains to be positive throughout the evolution of the Universe.
We then divide \eqref{consistency1_alphaA} by $M^2\dot{H}$ and solve \eqref{consistency1_alphaA} for the combination~$\alpha_B^2/\alpha_K$:
%
\begin{align}
\frac{\alpha_B^2}{\alpha_K} = -\frac{1}{12M^2\dot{H}}\left[  2M^2 \dot{H} + \sum_a \left(\bar{\rho}_a+\bar{p}_a \right)  \right]
\,. \label{eq:consistency}
\end{align}
%
The right-hand side is reminiscent of the background equation~\eqref{def_Fri2}. In particular, the choice~$M^2_{\rm cos}=M^2$ yields
%
\begin{align}
\frac{\alpha_B^2}{\alpha_K} = \frac{\bar{\rho}_{\rm DE}}{12M^2\dot{H}}(1+w_{\rm DE}) 
\qquad
({\rm for}~M^2_{\rm cos}=M^2)\,, \label{eq:consistency_wde}
\end{align}
%
and then
%
\begin{align}
w_{\rm DE}<-1\qquad
({\rm for}~M^2_{\rm cos}=M^2)\,.
\end{align}
%
Therefore, the DE has to be phantom-like (except for $\alpha_B=0$). The appearance of the phantom-like behaviour in the attractor phase of shift-symmetric scalar-tensor theories and in vector-tensor theories has been known in concrete models, see e.g.~\cite{DeFelice:2010pv, DeFelice:2016yws}. We, however, emphasise that this is a generic feature independent of the details of the models and is one of the consequences of the shift symmetry.

We should recall that the definition of $w_{\rm DE}$ (and $M^2_{\rm cos}$) is not unique because of the ambiguity of the notion of ``the DE component''. The above conclusion is obtained through the identification~$M^2_{\rm cos}=M^2$ under which the physical consequence of the shift symmetry is clear. On the other hand, in the code implementation, it would be useful to set $M^2_{\rm cos}=M_{\rm Pl}^2$ so that the background equations can be easily solved. 

The second consistency relation~\eqref{consistency2_alphaA} can be regarded as a differential equation of $M^2(t)$; we can solve \eqref{consistency2_alphaA} for $M^2(t)$ with given $\alpha$-parameters and the initial condition~$M^2(t_0)=M_{\rm Pl}^2$ at the present time. If $\alpha_T=\alpha_H=0$ is assumed (i.e.,~no deviation of the speed of GWs and no graviton decay~\cite{Creminelli:2018xsv}\footnote{Precisely speaking, the graviton decay constraint was obtained only in scalar-tensor theories and has not been explored in vector-tensor theories. Nonetheless, we can expect that the same three-point couplings among gravitons and longitudinal modes of the vector field would be generated by $\alpha_H\neq 0$. It would be interesting to explicitly discuss the graviton decay in vector-tensor theories and put constraints on $\alpha_H$ in a universal way.}), the solution is simply given by $M^2=M_{\rm Pl}^2$.

Having understood the physical implications, let us explain our code implementation of the consistency relations.
As we have mentioned in Sec.~\ref{sec:dictionary}, these relations trivially hold when the mapping between the concrete theory and the EFT parameters~\eqref{M2_H}--\eqref{alphaT_H} or \eqref{M2_GP}--\eqref{alphaV_GP} are used. We do not need to consider the consistency relations in the concrete model approach where the inputs are the functions of the Horndeski/generalised Proca Lagrangian. On the other hand, they have to be taken care of in the pure EFT approach. 
The simplest way to implement the consistency relations in the calculations is solving the equations~\eqref{consistency1_alphaA} and \eqref{consistency2_alphaA} for the EFT parameters. In our code implementation, we solve the first consistency relation~\eqref{consistency1_alphaA} for $\alpha_K$ with $M^2_{\rm cos}=M_{\rm Pl}^2$. Then, $\alpha_K$ is no longer an independent input of the EFT under the consistency relation while we need to confirm that the resultant solution of $\alpha_K$ must satisfy $\alpha_K>0$ for given inputs of other EFT parameters. This would be particularly useful because the sub-horizon physics may not be so much affected by $\alpha_K$; for instance, the effective gravitational coupling~\eqref{mu:H=0} and the slip parameter~\eqref{eta:H=0} are independent of $\alpha_K$ (note that $\alpha_H=0$ has been imposed in these expressions). The first consistency relation~\eqref{consistency1_alphaA} would not significantly affect the sub-horizon physics. We will study the effect on the CMB spectrum on large scales in Sec.~\ref{sec:shift_symmetry}. The second consistency relation~\eqref{consistency2_alphaA} is used to determine the dynamics of $M^2(t)$. In the present paper, however, we shall restrict our attention to the class with $\alpha_T=\alpha_H=0$ in which $M^2=M_{\rm Pl}^2$ in shift-symmetric theories. In combination with the first consistency relation~\eqref{consistency1_alphaA}, the equation-of-state parameter has to satisfy $w_{\rm DE}<-1$ then.

\section{Demonstrations}
\label{sec:demonstrations}

We now numerically solve the evolution equations and demonstrate the effect of $\alpha_{V}$ and the consistency relations on the angular power spectrum.

\subsection{\texorpdfstring{EFT with $\alpha_V$}{EFT with alphaV}}

As discussed in Sec.~\ref{sec:consistency}, the $\Lambda$CDM background, corresponding to $w_{\rm DE}=-1$, is prohibited in shift-symmetric theories with $\alpha_B \neq 0$ including vector-tensor theories. Hence, we should consider the $w$CDM model with $w_{\rm DE}<-1$ to model the corresponding background evolution.
We simply assume that $\alpha_{T}=\alpha_{H}=\alpha_{M}=\alpha^{GC}_{B}=\alpha^{GC}_{M}=\alpha^{GLPV}_{B}=0$ and vary $\alpha_{V,0}$, $\alpha_{B,0}$, and $w_{\rm DE}$. The remaining parameter, $\alpha_K$, is determined from the consistency relation in Eq.~(\ref{eq:consistency_wde}), whereas the other consistency relation~(\ref{eq:consistency}) is automatically satisfied. 
The evolution equations~(\ref{eq:pert2}) and (\ref{eq:pert3}) are continuously reduced to those in GR with the $w$CDM background if we take the limit where $\alpha_B, \alpha_K, \alpha_V \to 0$. We implement these equations to the Boltzmann solver developed by one of the authors in Refs.~\cite{Hiramatsu:2020fcd,Hiramatsu:2022fgn}. In the code, we use the conformal time, $\D\eta = \D t/a$ (which is not to be confused with the slip parameter).

\begin{figure}[t]
    \centering{
    \includegraphics[width=12cm]{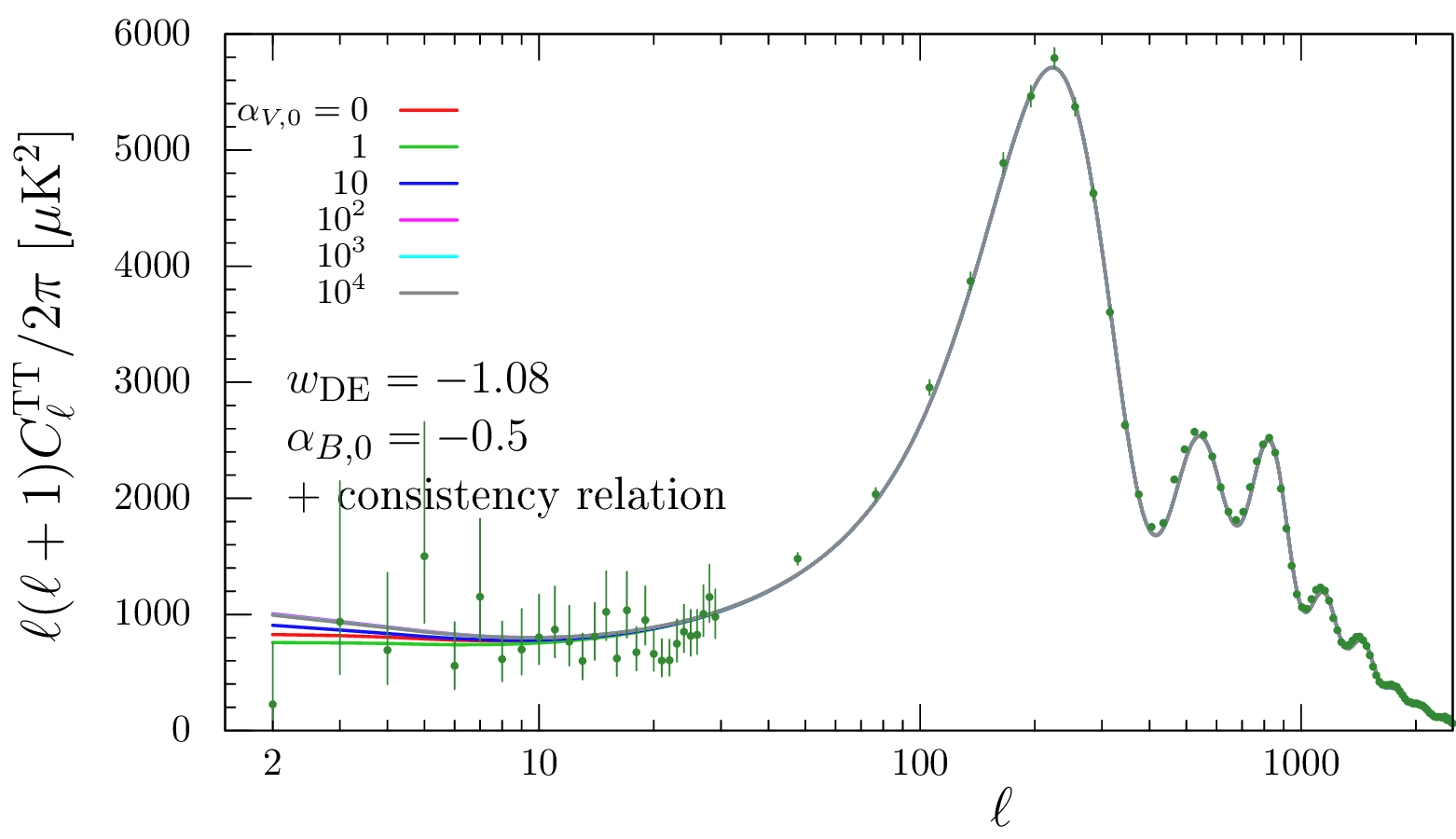}
    }
    \caption{Angular power spectrum of the temperature fluctuations with $\alpha_{B,0}=-0.5$ and $w_{\rm DE}=-1.08$. }
    \label{fig:demonstration1_whole}
\end{figure}

As a demonstration, we assume a simple time dependence of the $\alpha$-parameters,
%
\begin{equation}
    \alpha_{i}(\eta) \equiv \frac{\Omega_{\rm DE}(\eta)}{\Omega_{\rm DE, 0}} \alpha_{i,0}\,,
\label{alpha_parameterisation}
\end{equation}
%
where $\alpha_{i,0}$ is the present value of $\alpha_i(\eta)$ and $\Omega_{\rm DE} \equiv 1-\Omega_{\rm m}-\Omega_{\rm r}$.
In Fig.~\ref{fig:demonstration1_whole}, we show the rescaled angular power spectrum of the temperature fluctuations, $C^{\rm TT}_{\ell}$, with $\alpha_{B,0}=-0.5$ and $w_{\rm DE}=-1.08$, whereas $\alpha_{K,0}$ is determined from the consistency relation~(\ref{consistency1_alphaA}). The green points with error bars represent the binned data from Planck 2018 results~\cite{Planck:2018vyg}. This figure shows the dependence of the angular power spectrum on $\alpha_{V,0}$ characterising the gauge coupling. 
In our present setup, the extra degree of freedom mediating gravity plays a role of DE, so that its impact on the (unlensed) angular power spectrum appears only on the large scales, typically $\ell \lesssim \mathcal{O}(10)$. Hence, we hereafter show only the spectrum for $\ell < 100$. In this demonstration, $w_{\rm DE}$ is shifted significantly from $-1$ (cosmological constant) to magnify the differences in the spectrum due to the effect of $\alpha_{V,0}$ and we do not take care of the consistency to the observational result.

\begin{figure}[t]
    \centering{
    \includegraphics[width=8cm]{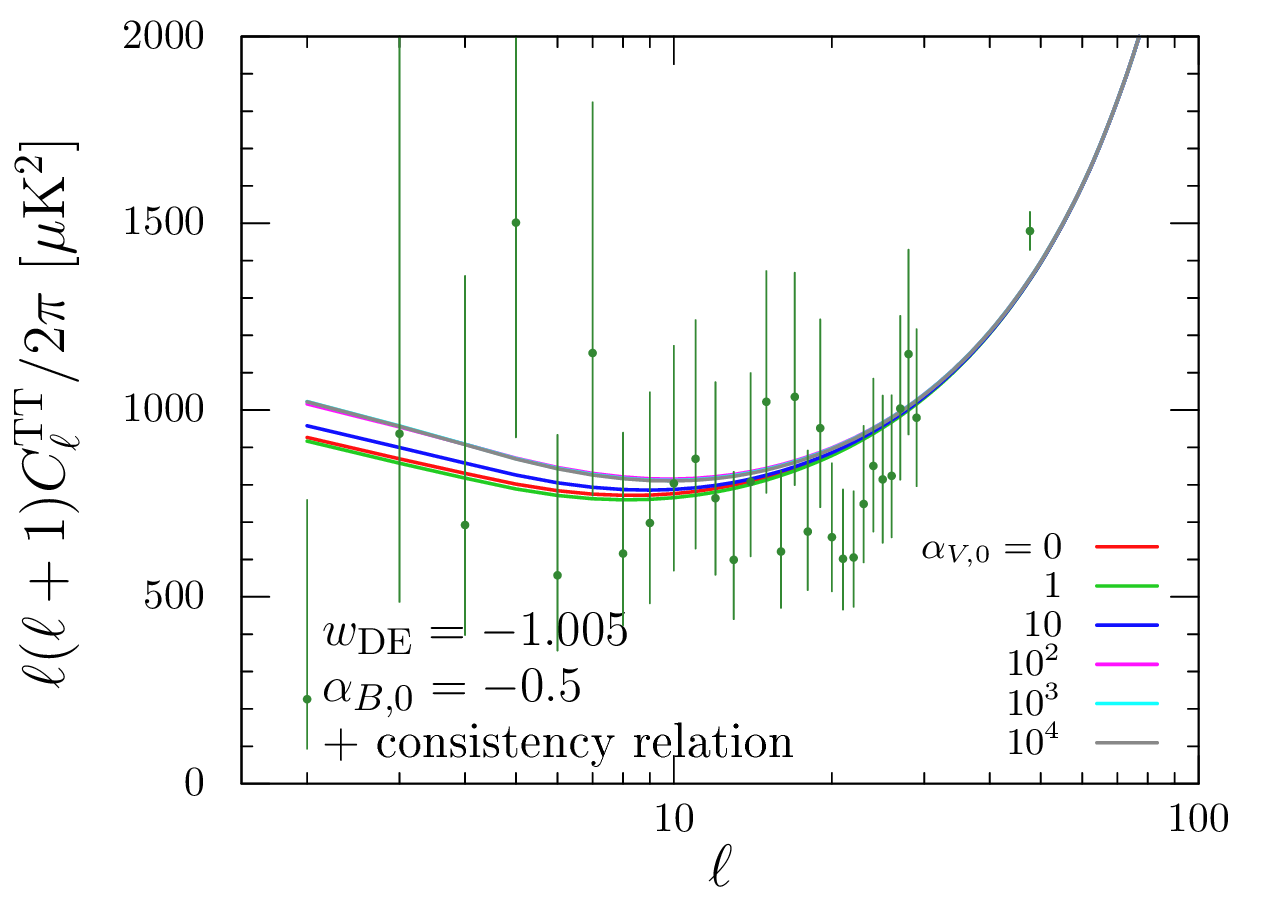}
    \includegraphics[width=8cm]{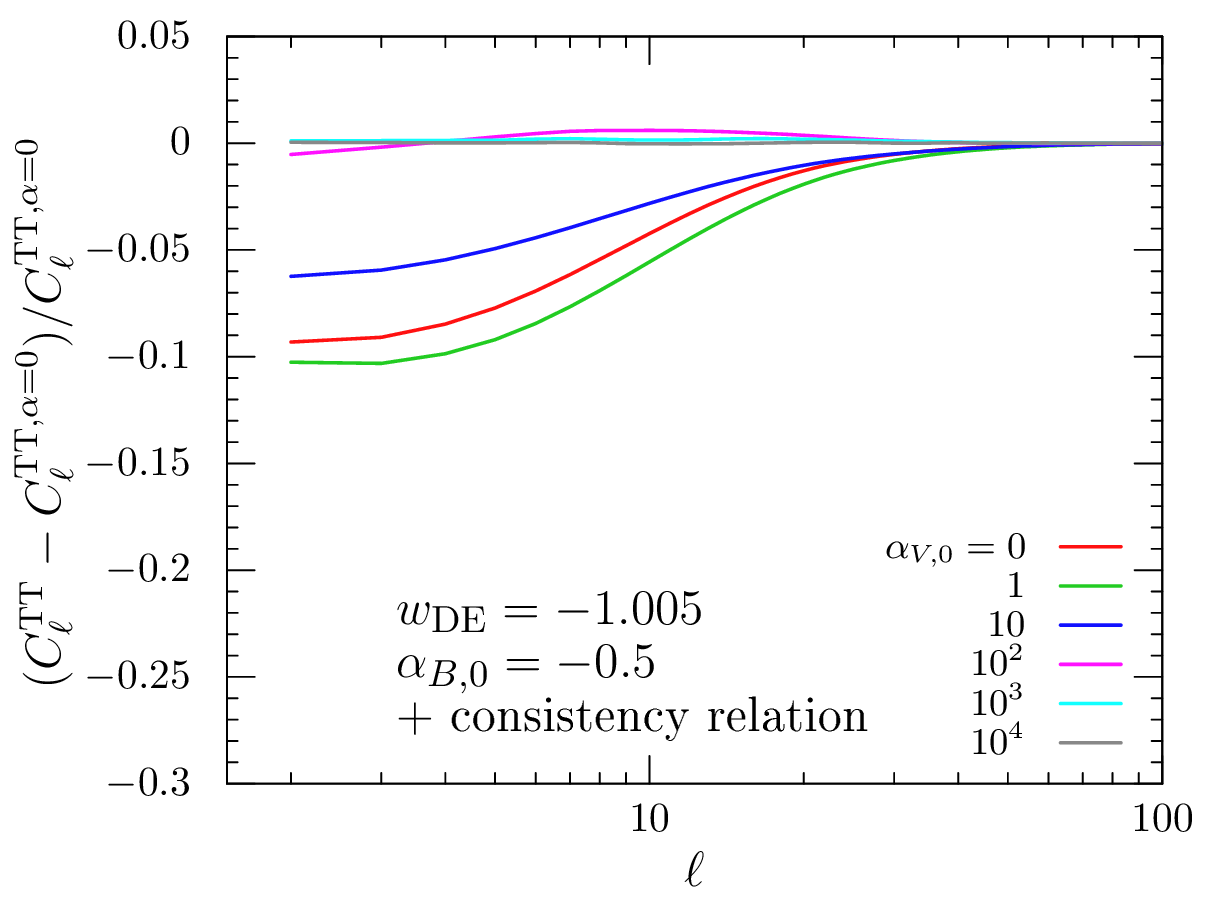}
    }
    \centering{
    \includegraphics[width=8cm]{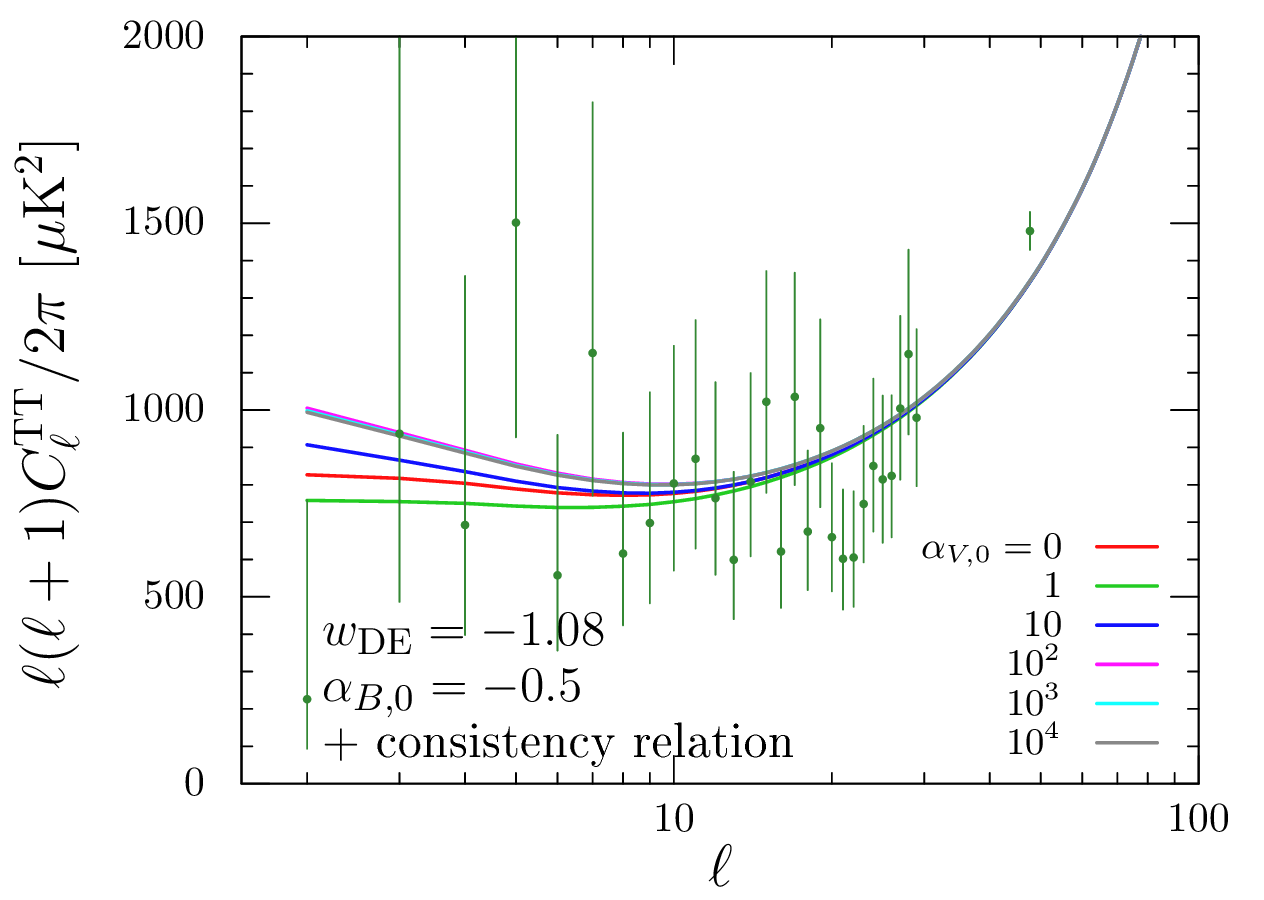}
    \includegraphics[width=8cm]{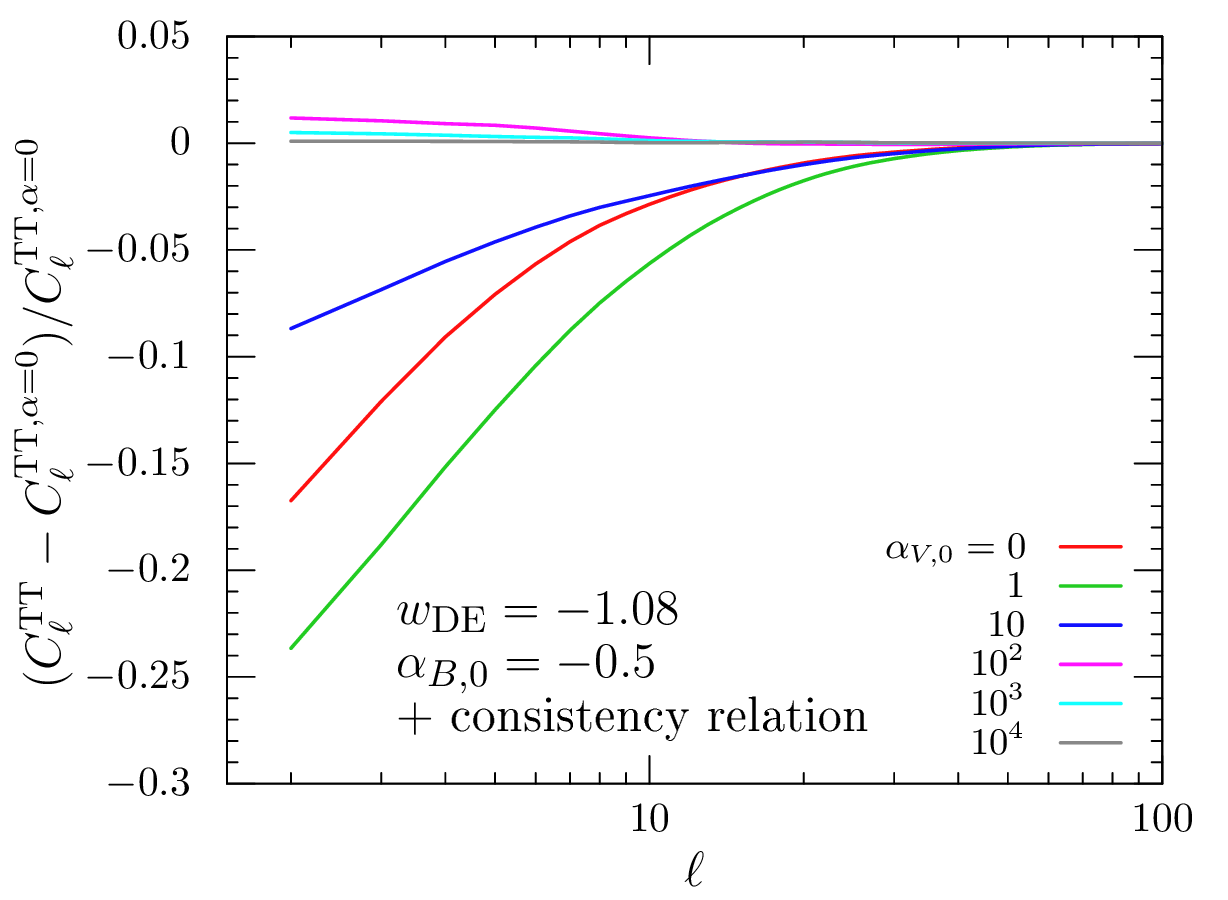}
    }
    \caption{Angular power spectrum of the temperature fluctuations with $\alpha_{B,0}=-0.5$ and $w_{\rm DE}=-1.005$ (top) and $-1.08$ (bottom). The left panels show the temperature angular power spectrum for low multipoles and the green points with error bars represent the observational results from Planck 2018~\cite{Planck:2018vyg}. The right panels show the relative errors to the temperature angular power spectra in the limit~$\alpha_K=\alpha_B=\alpha_V=0$.}
    \label{fig:demonstration1}
\end{figure}

In Fig.~\ref{fig:demonstration1}, we show the cases with $\alpha_{B,0}=-0.5$ and $w_{\rm DE}=-1.005$ (top) and $-1.08$ (bottom) for $\ell \leq 100$. The right panels show the relative error to the case with $\alpha_i=0$ for $i=K,B,V$ (no modification of gravity).\footnote{We cannot directly take the limit, $\alpha_i\to 0$, in Eq.~(\ref{eq:pert1}) since the equation becomes singular. Instead, we solve Eqs.~(\ref{eq:pert2}) and (\ref{eq:pert3}) with $\Pi=0$ in this limit.} If $\alpha_{V}=0$, the case of scalar-tensor theories is recovered, in which the extra scalar degree of freedom highly suppresses the CMB anisotropy by $\sim 20\%$ on the low multipoles as shown in the red curve. This feature is consistent with the previous studies~\cite{DAmico:2016ntq,Hiramatsu:2020fcd}. Then, increasing $\alpha_{V,0}$, we find that the anisotropy is less suppressed and eventually the spectrum recovers the one in GR (i.e.,~$w$CDM background with $\alpha_i=0$). This feature is expected in Eq.~(\ref{alpha_relation}).
From Eq.~(\ref{eq:def_G}), we find that $\mathcal{G}\approx 1$ for $k \ll aH\sqrt{\alpha_V\alpha_K}$, leading to $\tilde{\alpha}_B,\tilde{\alpha}_K \to 0$. Hence, large $\alpha_V$ suppresses the effects of the scalar-tensor theories on large scales. This feature appears even with large $|w_{\rm DE}|$ (bottom panels).

In scalar-tensor theories, the angular power spectrum is changed from that in GR due to the extra scalar degree of freedom. Hence, such theories can possibly be constrained from the CMB observations as long as the deviation is larger than the cosmic variance. In contrast, in the vector-tensor theories, the vector field can erase such effects. This characteristic property can potentially evade the observational constraints of the large-scale survey of the CMB.

\subsection{EFT with and without consistency relations}
\label{sec:shift_symmetry}

Next, we explore the impact of the consistency relation~(\ref{consistency1_alphaA}) on the CMB spectrum. The consistency relations are the requirement of the shift symmetry and can be violated in scalar-tensor theories when, e.g., the scalar field has a potential. We compare the spectra with the consistency relation where $\alpha_K$ is determined by (\ref{consistency1_alphaA}) and without it where $\alpha_K$ is determined by the parameterisation~\eqref{alpha_parameterisation}. Since vector-tensor theories must satisfy the consistency relation, we only focus on scalar-tensor theories $(\alpha_V=0)$ in this subsection.

\begin{figure}[t]
    \centering{
    \includegraphics[width=8cm]{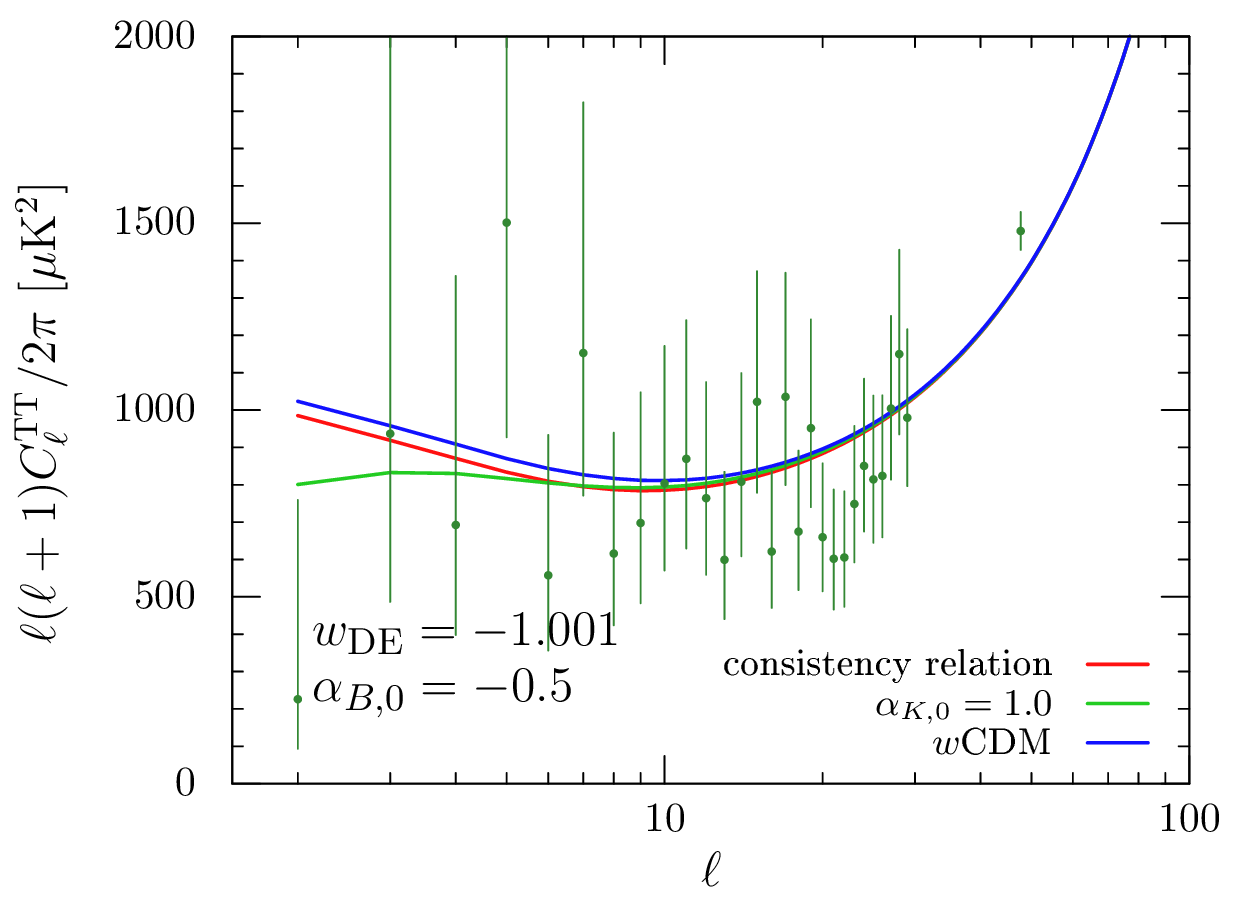}
    \includegraphics[width=8cm]{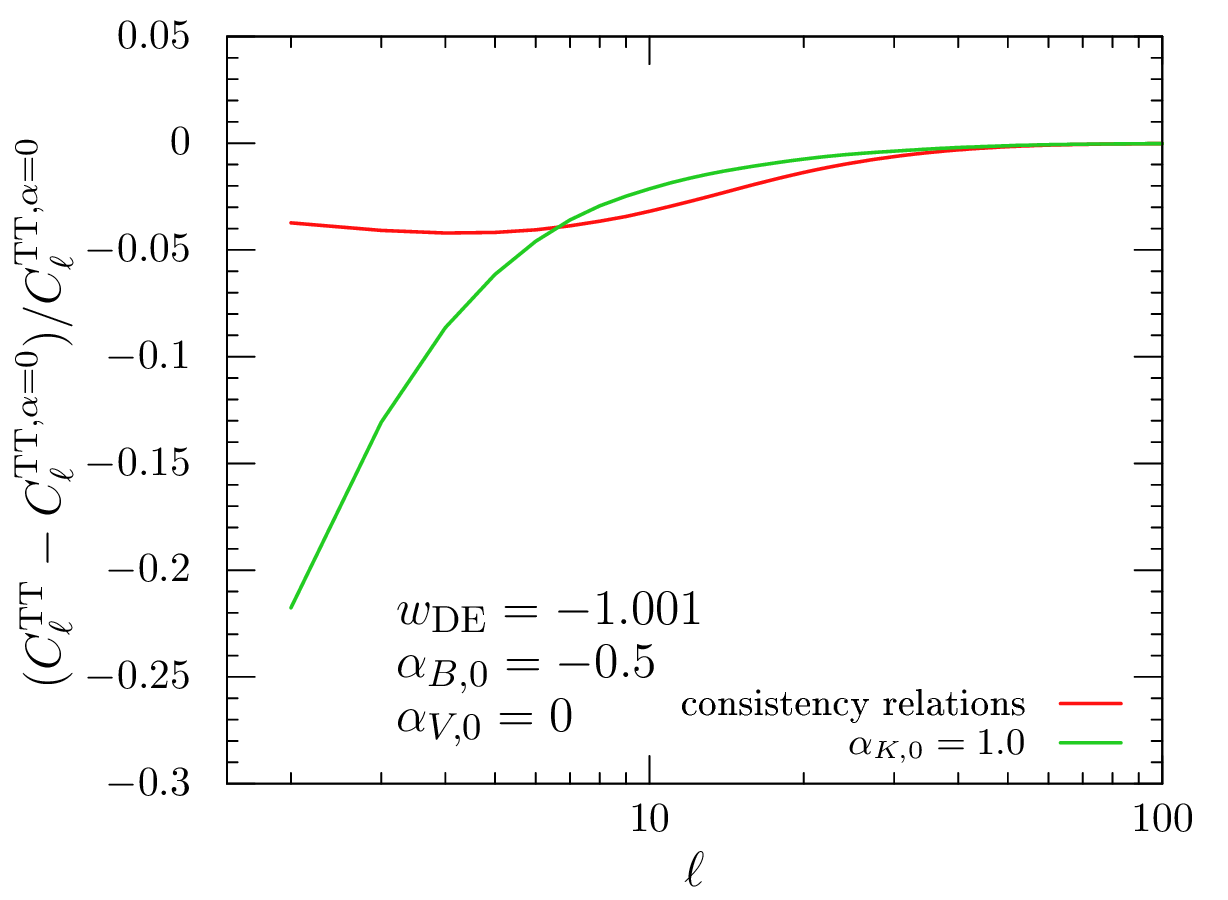}
    }
    \centering{
    \includegraphics[width=8cm]{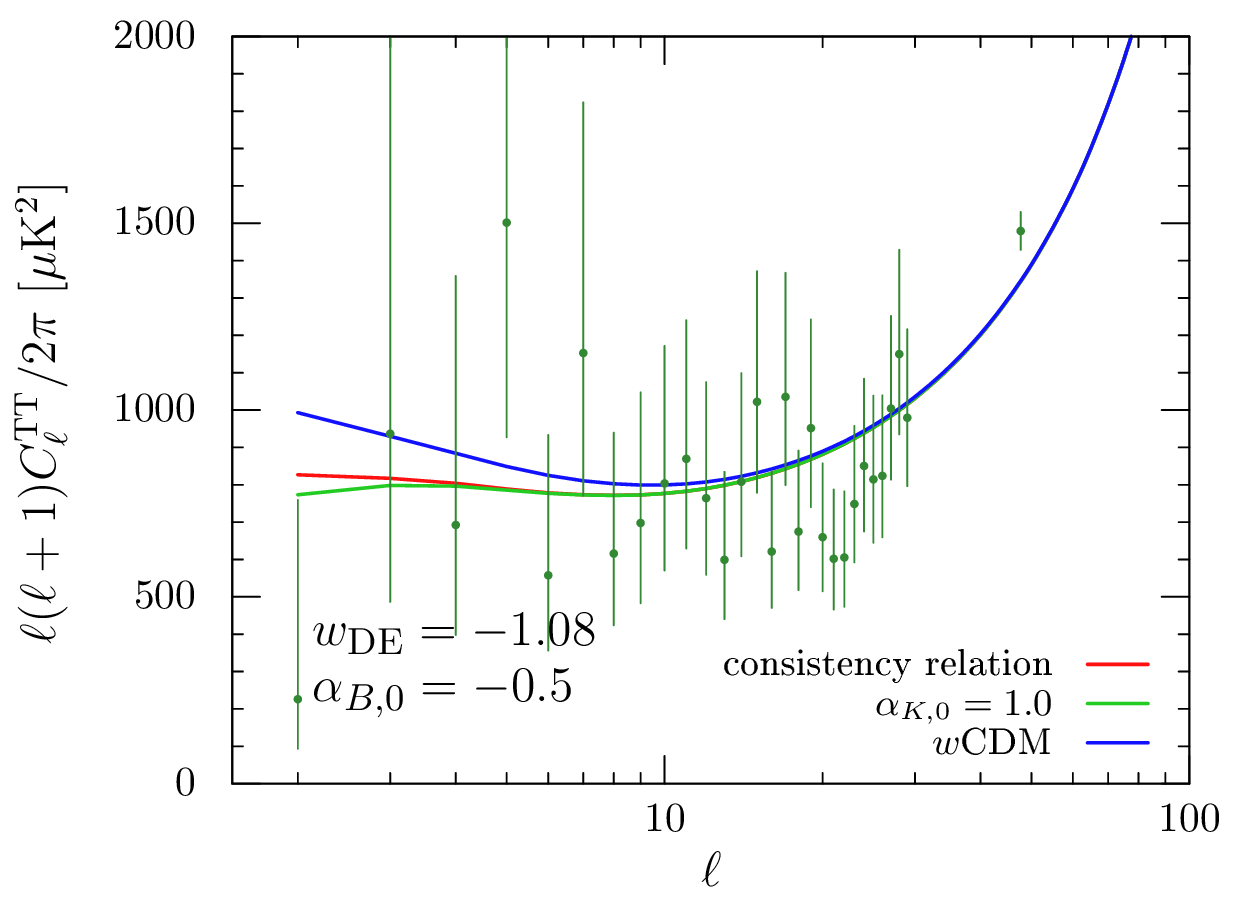}
    \includegraphics[width=8cm]{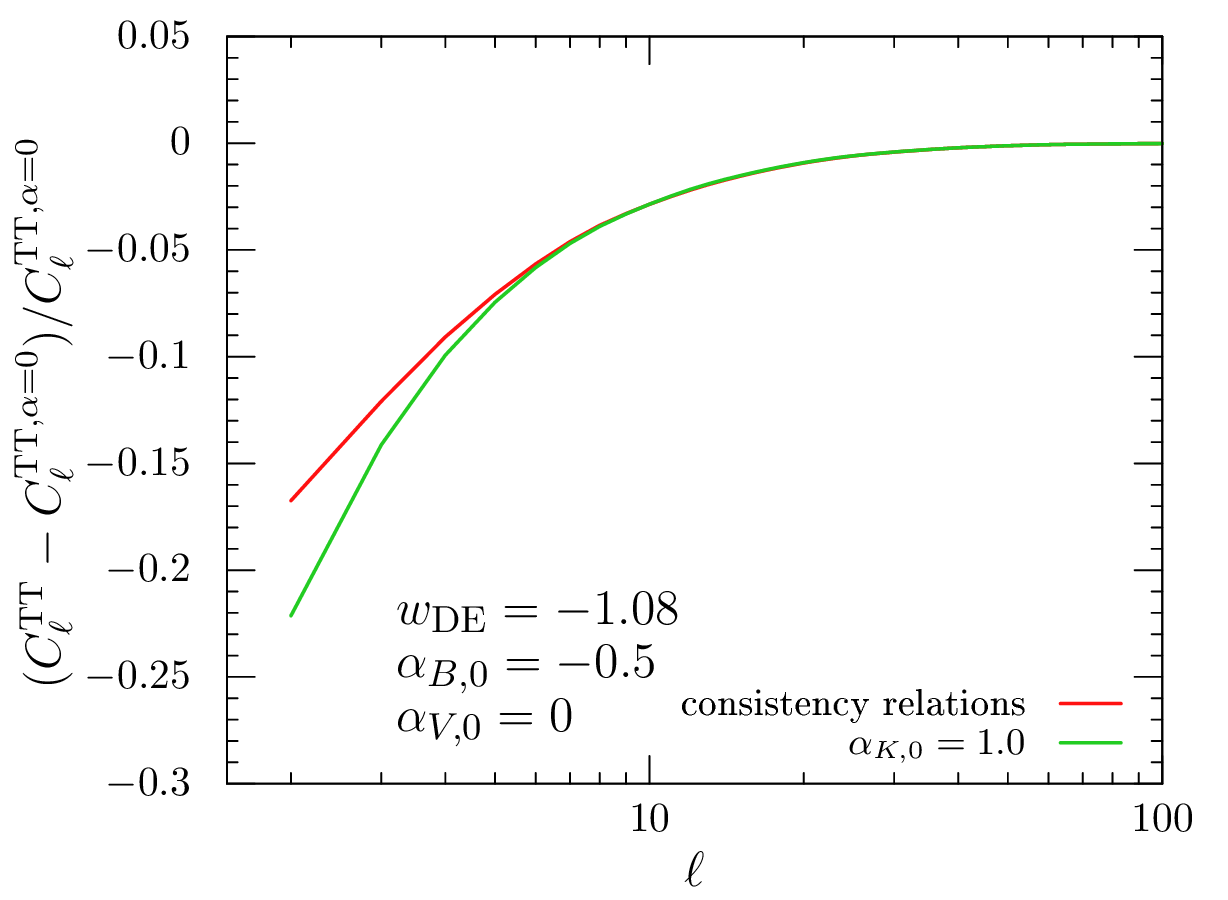}
    }
    \caption{The same figure as Fig.~\ref{fig:demonstration1} for the case with the consistency relations (red), without the consistency relations (green), and the $w$CDM model (blue).
}
    \label{fig:demonstration0}
\end{figure}

In Fig.~\ref{fig:demonstration0}, we show the angular power spectra with $\alpha_{B,0}=-0.5$ and $w_{\rm DE}=-1.001$ (left) and $-1.08$ (right) in the $w$CDM background. The case where $\alpha_{K}$ is determined from the consistency relation is shown in red lines, while the case with fixed $\alpha_{K,0}$ is in green lines. The blue lines show the angular power spectrum with $\alpha_i=0$ as a reference. In the right panels, we show the relative errors to the blue lines in both cases. If $w_{\rm DE}=-1.08$, the two cases with and without the consistency relations are not so different from each other. In contrast, if $w_{\rm DE}$ is close to the cosmological constant, $w_{\rm DE}=-1.001$, the difference between with and without the consistency relations is more significant than the case with the large $|w_{\rm DE}|$. It is easily understood by \eqref{eq:consistency_wde}. For $w_{\rm DE}\approx -1$, the consistency relation~\eqref{eq:consistency_wde} leads to large $\alpha_K$ for a fixed $\alpha_B$, since the right-hand side, which is proportional to $(1+w_{\rm DE})\rho_{\rm DE}$, becomes close to zero. There should be a large kineticity effect to compensate $w_{\rm DE} \to -1$. 

Hence, we conclude that the consistency relation~\eqref{consistency1_alphaA} does not so much affect perturbations except for near $\Lambda$CDM backgrounds. The crucial role of \eqref{consistency1_alphaA}, namely the shift symmetry, would rather be a restriction of the background dynamics so that $w_{\rm DE}<-1$.

\subsection{Concrete model}

Finally, we compute the angular power spectrum based on a concrete model of DE described by the following subclass of generalised Proca theories:
%
\begin{align}
\mathcal{L} = \frac{M_{\rm Pl}^2}{2}R -\frac{1}{4g^2}F_{\mu\nu}F^{\mu\nu} + g_2(X) + g_3(X) \nabla_{\mu} A^{\mu}
\,,\qquad X=-\frac{1}{2}A_{\mu}A^{\mu}
\,,
\label{concrete_model}
\end{align}
%
where the functions are chosen as power-law forms~\cite{DeFelice:2016yws, deFelice:2017paw, Nakamura:2018oyy, DeFelice:2020sdq},
%
\begin{align}
g_2=b_2 X^{p_2}\,, \qquad g_3 = b_3 X^{p_3}
\,,
\label{g2g3}
\end{align}
%
with $g$ being the gauge coupling and $b_2$, $p_2$, $b_3$, $p_3$ being constants.
The scalar-tensor limit corresponds to $g\to 0$ after the replacement~$A_{\mu} \to g A_{\mu}+\partial_{\mu}\phi$. We then reach the Horndeski counterpart of \eqref{concrete_model}:
\begin{align}
\mathcal{L} = \frac{M_{\rm Pl}^2}{2}R  + g_2(X) + g_3(X) \nabla_{\mu} \nabla^{\mu} \phi
\,,\qquad X=-\frac{1}{2}\nabla_{\mu}\phi \nabla^{\mu}\phi
\,,
\label{concrete_model2}
\end{align}
with the same functional forms as \eqref{g2g3}.

The EFT functions are obtained by using the dictionary presented in Sec.~\ref{sec:dictionary}. The constraint equation on the vector field or the attractor phase of the scalar field, $J=0$, leads to
%
\begin{align}
H=\frac{b_2 p_2}{3\sqrt{2}\,b_3 p_3}X^{p_2-p_3-1/2}\,, \label{eq:constraint_H}
\end{align}
%
by which we can solve $X$ for the Hubble expansion rate~$H$. Keeping $X$ for notational convenience, the effective energy density and pressure of DE are
%
\begin{align}
    \bar{\rho}_{\rm DE}=-b_2 X^{p_2}\,, \qquad \bar{p}_{\rm DE}=b_2 X^{p_2}\left(1+\frac{p_2\dot{X}}{3HX} \right) \,,
    \label{eq:def_rhoDE}
\end{align}
%
while the $\alpha$-parameters are found as
%
\begin{align}
    M^2&=M_{\rm Pl}^2 \,, \\
    \alpha_K &=-\frac{2b_2 p_2(2p_3-2p_2+1) X^{p_2}}{M_{\rm Pl}^2H^2} = 6p_2(2p_3-2p_2+1) \Omega_{\rm DE} \,, \\
    \alpha_B &=\frac{b_2p_2 X^{p_2}}{3M_{\rm Pl}^2 H^2} =-p_2 \Omega_{\rm DE} \,, \\
    \alpha_T &=0\,, \\
    \alpha_V &=\frac{g^2 M_{\rm Pl}^2}{2X}\,, \label{eq:proca_alphaV}
\end{align}
%
with $\Omega_{\rm DE}=\bar{\rho}_{\rm DE}/(3M_{\rm Pl}^2H^2)$ being the density parameter of DE.

The background is obtained by solving the Friedmann equation~$H^2 = (\bar{\rho}_{\rm m}+\bar{\rho}_{\rm r}+\bar{\rho}_{\rm DE})/(3M_{\rm Pl}^2)$ and the constraint equation~(\ref{eq:constraint_H}), where $\bar{\rho}_{\rm m}$ and $\bar{\rho}_{\rm r}$ are the energy density of the non-relativistic matter and the radiation, respectively, and $\bar{\rho}_{\rm DE}$ is given in Eq.~(\ref{eq:def_rhoDE}).

\begin{figure}[t]
    \centering{
    \includegraphics[width=8cm]{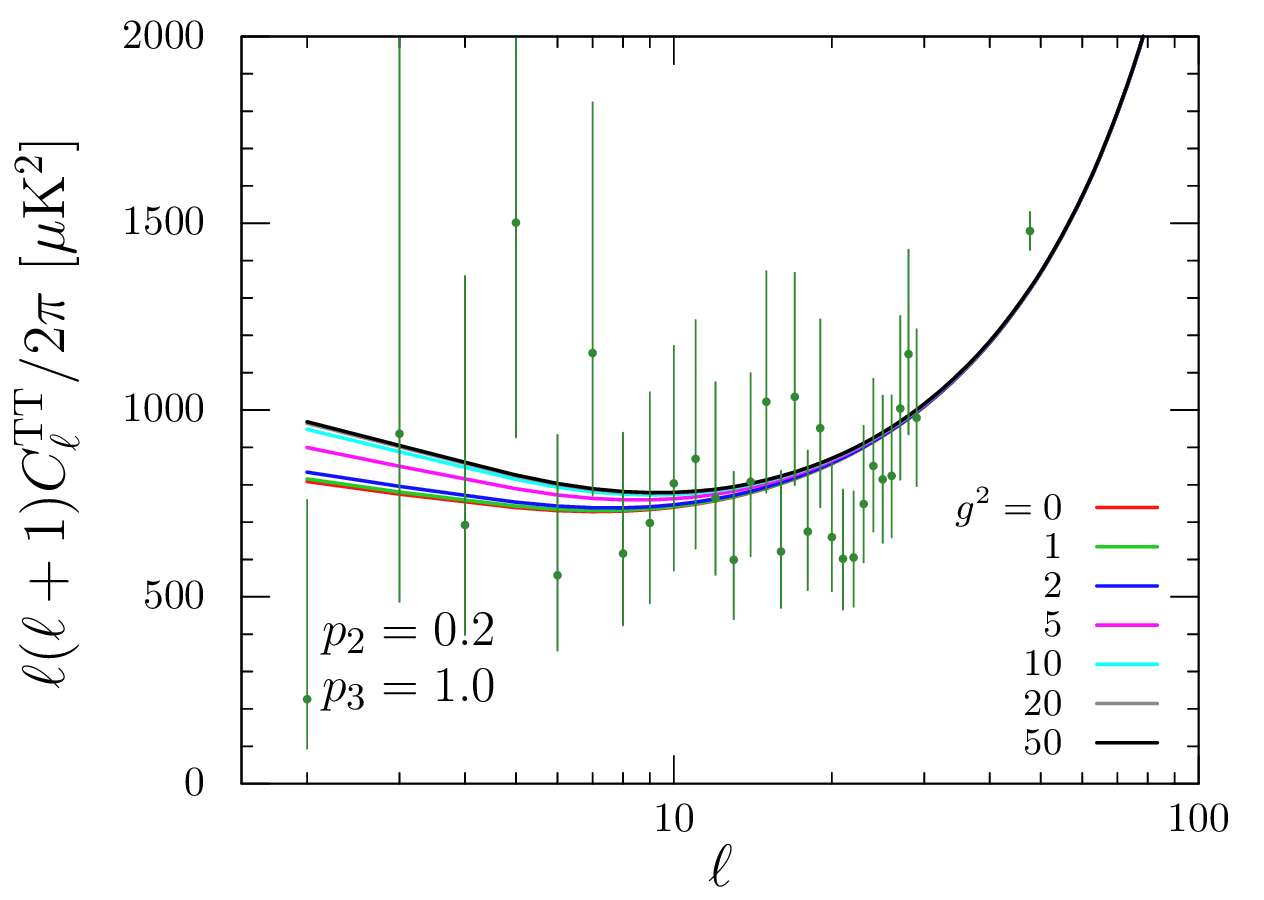}
    \includegraphics[width=8cm]{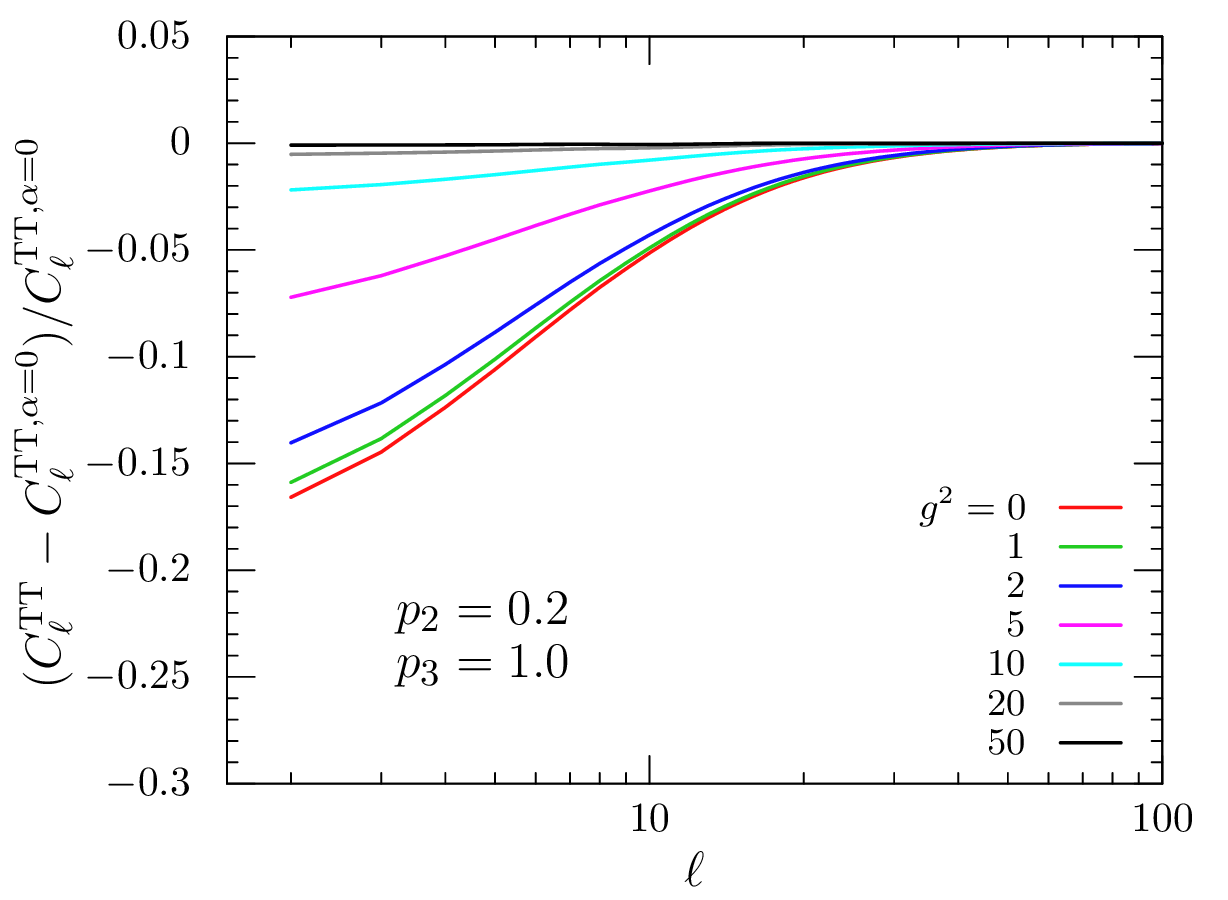}
    }
    \caption{Temperature angular power spectrum in a power-law model as a subclass of the generalised Proca theory (left) and the relative error to that in GR with the same cosmic expansion history as in the power-law model (right).}
    \label{fig:demonstration2}
\end{figure}

We show the angular power spectrum of the temperature anisotropy with $p_2=0.2$ and $p_3=1.0$ as a demonstration in Fig.~\ref{fig:demonstration2}. In the right panel, we show the relative error to the case with $\alpha_i=0$ keeping the same background dynamics. We vary the gauge coupling constant~$g$, which is related to $\alpha_V$ parameter as in Eq.~(\ref{eq:proca_alphaV}). The case~$g=0$ corresponds to the Horndeski theory with the same power-law functions. As in the previous demonstration, the large gauge coupling or large $\alpha_V$ suppresses the deviation from GR.

\section{Conclusions} \label{sec:conclusion}
There are a plethora of modified gravity theories that have been developed until now. If we take the subclass of models known as scalar-tensor theories, there are a number of features that can have a phenomenological impact due to terms in the Lagrangian. All these models within the scalar field DE framework are unified with an effective description to accommodate all possible phenomena. This unified description is usually referred to as EFT of DE. However, this description misses another interesting class of modified gravity theories, namely, vector-tensor theories of gravity. Recently, there has been a development in describing both scalar-tensor and vector-tensor theories in a unified framework, which can accommodate all the features of these models~\cite{Aoki:2021wew}. The new ingredients are a new EFT coefficient function~$\alpha_V(t)$ and the set of consistency relations which universally characterise the vectorial nature of DE and the presence/absence of the shift symmetry, respectively.

In this work, we have explored the impact of this unified EFT of DE in the CMB power spectrum. We have focused on the class of Horndeski/generalised Proca theories with the luminal speed of GWs and found that the new ingredients of the EFT have simple implications: the vectorial nature suppresses modifications of gravity and the shift symmetry implies phantom DE. The parameter~$\alpha_V$ connects the scalar-tensor cosmology $(\alpha_V=0)$ to the $w$CDM cosmology $(\alpha_V \to \infty)$ with the same background dynamics as demonstrated in Fig.~\ref{fig:demonstration1}. A similar conclusion has been reached in the concrete model (Fig~.\ref{fig:demonstration2}). On the other hand, as shown in Fig.~\ref{fig:demonstration0}, the shift symmetry does not have much influence on the CMB spectrum for a given background dynamics unless the background is close to the $\Lambda$CDM. Instead, the shift symmetry, either global (scalar-tensor) or local (vector-tensor), restricts the background dynamics so that $w_{\rm DE}<-1$.

In conclusion, vector-tensor theories can prevent the modification of gravity while keeping the same background dynamics as in scalar-tensor theories which can be used to evade observational constraints of modified gravity on large scales. In particular, the large-$\alpha_V$ limit gives a phantom DE cosmology. On the other hand, if $w_{\rm DE}>-1$ is observationally verified, which is indeed preferred in the latest result of the DESI+CMB combination~\cite{DESI:2024mwx}, vector-tensor theories and shift-symmetric scalar-tensor theories in the attractor phase may be excluded as $w_{\rm DE}>-1$ tends to be incompatible with the shift symmetry. A precise analysis is required to conclude the observational viability of these theories. It would also be interesting to investigate universal predictions in other scales such as large-scale structures and astrophysical scales.\footnote{Along this line of thought, EFTs for black hole perturbations were developed in~\cite{Mukohyama:2022enj, Khoury:2022zor, Mukohyama:2022skk, Mukohyama:2023xyf, Konoplya:2023ppx, Aoki:2023bmz}.} We leave them for future works.

\acknowledgements
The work of K.A.~was supported by JSPS KAKENHI Grant Nos.\ JP20K14468 and JP24K17046.
The work of M.A.G.~was supported by Mar\'{i}a Zambrano fellowship. 
The work of T.H.~was supported by JSPS KAKENHI Grant Nos.\ JP21K03559 and JP23H00110.
The work of S.M.~was supported in part by JSPS KAKENHI Grant No.\ JP24K07017 and the World Premier International Research Center Initiative (WPI), MEXT, Japan.
M.C.P.~acknowledges the Mahidol University International Postdoctoral Fellowship. M.C.P.~is also partially supported by Fundamental Fund: fiscal year 2024 by National Science Research and Innovation Fund (NSRF).
The work of K.T.~was supported by JSPS KAKENHI Grant No.\ JP23K13101.

\appendix*
\section{Unified EFT of scalar-tensor and vector-tensor theories in cosmology}
\label{sec:matter}

In cosmology, a dynamical background vector field generates anisotropy and that is why most of the studies are restricted to the simple case of the scalar field. The so-called EFT of vector-tensor cosmology \cite{Aoki:2021wew} is based on the fact that even non-dynamical isotropic {\it background} vector field provides a non-trivial phenomenology. In this appendix, we briefly review the scalar sector of the EFT of vector-tensor cosmology in which we are interested in this paper.

\subsection{Fully nonlinear action}

We assume that the spacetime symmetry is spontaneously broken by the presence of a timelike vector field which in unitary gauge takes the form
\begin{align}\label{vector-deltaT}
\tilde{\delta}^0_\mu = {\delta}^0_\mu + \g A_\mu \,,
\end{align}
where $\g$ is the gauge coupling constant. The above vector is invariant under the combination of the time diffeomorphisms and the $U(1)$ gauge symmetry which is the residual symmetry of the EFT on top of the time-dependent spatial diffeomorphisms. The latter is the only residual symmetry of the EFT of the scalar-tensor theories while the former downgrades to the shift symmetry of the scalar-tensor theories if $\g=0$ and if we ignore the transverse degrees of freedom in $A_\mu$~\cite{Aoki:2021wew}. Let us show these facts more explicitly. We construct the unit future-directed timelike vector out of \eqref{vector-deltaT} as follows
\begin{align}\label{vector-nT}
&\tilde{n}_\mu = - \frac{\tilde{\delta}^0_\mu}{\sqrt{-\tilde{g}^{00} }} \,; \qquad
\tilde{g}^{00} 
\equiv \tilde{\delta}^0_\alpha \tilde{\delta}^0_\beta g^{\alpha\beta} \,.
\end{align}
Ignoring the transverse degrees of freedom (focusing on the irrotational solutions), we have $A_\mu=(A_0,\partial_iA)$. Using the residual gauge freedom of the combined $U(1)$ and time diffeomorphisms, we can choose the gauge such that $A=0$, giving
\begin{align}\label{A-A0}
A_\mu = A_0 \delta^0_\mu \,.
\end{align}
For the above temporal vector field, the norm of the preferred vector~\eqref{vector-deltaT} is
\begin{align}\label{tildeg00_ST}
\tilde{g}^{00}=g^{00}(1+\g A_0)^2 \,,
\end{align}
and we find 
\begin{align}\label{vector-n}
&\tilde{n}_\mu = - \frac{{\delta}^0_\mu}{\sqrt{-{g}^{00} }} = n_\mu \,,
\end{align}
where $n_\mu$ is the normal vector to the constant-time hypersurfaces. Thus, in the absence of the transverse degrees of freedom, the vector~$\tilde{n}_\mu$ coincides with $n_\mu$ which is the building block of the EFT of scalar-tensor theories in the unitary gauge when $\phi=t$. Note that still the presence of the gauge coupling~$\g$ and also the extra residual symmetry (combination of the time diffeomorphisms and the $U(1)$ gauge symmetry) make the underlying EFT different than the usual EFT of the scalar-tensor theories (see Fig.~1 in \cite{Aoki:2021wew}).
Having the normal vector~$n_\mu$ in hand, we can define the induced metric on the constant-time hypersurfaces as
\begin{align}
h_{\mu\nu} \equiv g_{\mu\nu} + n_\mu n_\nu \,,
\end{align}
and the corresponding extrinsic curvature as
\begin{align}
K_{\mu\nu} \equiv h_{(\mu|}{}^\alpha\nabla_\alpha n_{|\nu)} \,.
\end{align}

At the background level, the metric is given by \eqref{FLRW}. Decomposing the vector field~\eqref{A-A0} into the background and perturbation parts as 
\begin{align}\label{A-ST} A_0=\bar{A}_0(t)+\delta A_0(t,x^i)
\,,
\end{align}
the EFT action of the perturbations in the unitary gauge  takes the following form~\cite{Aoki:2021wew}:
\begin{align}\label{action-ST}
S 
&=\int \D^4x \sqrt{-g} \Big[
\mathcal{L}_{\rm DE}^{(0)} 
+ \mathcal{L}_{\rm DE}^{(2)} 
\Big] + S^{\rm m} \,, 
\\
\mathcal{L}_{\rm DE}^{(0)} 
&=  \frac{\Mpl^2}{2}f(t) \left[ \spatialR +K_{\mu\nu}K^{\mu\nu}-K^2 \right] - \Lambda(t)-c(t) \tilde{g}^{00}-d(t)K \,,
\label{Lagrangian-0} 
\\ \nonumber
\mathcal{L}_{\rm DE}^{(2)} 
&= 
\frac{1}{2} M_2^4(t) \left(\frac{\delta\tilde{g}^{00}}{-\tilde{g}_{\rm BG}^{00} }\right)^2
- \frac{1}{2} {\bar M}_1^3(t) \left(\frac{\delta\tilde{g}^{00}}{-\tilde{g}_{\rm BG}^{00} }\right) \delta  K
- \frac{1}{2} {\bar M}_2^2(t) \delta K^2
- \frac{1}{2} {\bar M}_3^2(t) \delta K^{\alpha}{}_{\beta} \delta K^{\beta}{}_{\alpha} 
\\
&\quad + \frac{1}{2}\mu_1^2(t)\left(\frac{\delta\tilde{g}^{00}}{-\tilde{g}_{\rm BG}^{00} }\right) \delta \spatialR - \frac{1}{4} \gamma_1(t) F_{\alpha\beta} F^{\alpha\beta}  
+\frac{1}{2}\bar{M}_{4}(t) \delta K  \delta \spatialRST
+\frac{1}{2}\lambda_{1}(t) \delta \spatialR^2
+ \cdots\,,
\label{Lagrangian-2}
\end{align}
where the ellipsis stands for the higher-derivative terms and all the EFT coefficients are functions of time only. In the above action, $S^{\rm m}$ is the total matter action, $\spatialR$ is the spatial Ricci scalar, $\tilde{g}_{\rm BG}^{00}$ denotes the background value of $\tilde{g}^{00}$, $F_{\mu\nu}=\partial_\mu A_\nu-\partial_\nu A_\mu$ is the strength tensor for the gauge field, $K=g^{\alpha\beta}K_{\alpha\beta}$ is the trace of the extrinsic curvature, and 
\begin{align}\label{deltaK-def}
&\delta {K}^{\mu}{}_{\nu} = {K}^{\mu}{}_{\nu}-H(t)h^{\mu}{}_{\nu} \,, \qquad
\delta {K} = K - 3H(t) \,.
\end{align}
The background equations of motion are given by
\begin{align}
c(t)&=0 \,, \label{c_eq}\\
\Lambda(t)&=-\bar{\rho}_{\rm m}+3 \Mpl^2  f H^2
\,, \label{lambda_eq}
\\
\dot{d}(t)&= - \bar{\rho}_{\rm m} - \bar{p}_{\rm m} - 2\Mpl^2(  f \dot{H} + \dot{f} H )
\,, \label{d_eq} 
\end{align}
and from the conservation of the matter energy-momentum tensor, we have
\begin{align}
\dot{\bar{\rho}}_{\rm m} +3H (\bar{\rho}_{\rm m} +\bar{p}_{\rm m})=0 \,.
\label{conserve}
\end{align}

Let us now rewrite the Lagrangian~\eqref{Lagrangian-0} in terms of $\delta {K}^{\mu}{}_{\nu}$ and $\delta {K}$ that are defined in Eq.~\eqref{deltaK-def}. By the direct substitution, we find
\begin{eqnarray}
\mathcal{L}_{\rm DE}^{(0)} 
= \frac{\Mpl^2}{2}f \left[ \spatialR +\delta K^{\alpha}{}_{\beta} \delta K^{\beta}{}_{\alpha}-(\delta{K})^2 \right] 
+ 3 \Mpl^2 f H^2 - \Lambda - \left( 2 \Mpl^2 f H + d \right) K \,,
\end{eqnarray}
where we have used $c=0$ from the background equation~\eqref{c_eq}.
Performing an integration by parts with the relation~$K=\nabla_\mu n^\mu$, we find
\begin{eqnarray}
\mathcal{L}_{\rm DE}^{(0)} 
= \frac{\Mpl^2}{2}f \left[ \spatialR +\delta K^{\alpha}{}_{\beta} \delta K^{\beta}{}_{\alpha}-(\delta{K})^2 \right] 
+ 3 \Mpl^2 f H^2 - \Lambda + \frac{\bar{N}}{N} \left( 2 \Mpl^2 f H + d \right)^{\boldsymbol{\cdot}} \,,
\end{eqnarray}
where we have used $n^0 = \sqrt{-g^{00}}=N^{-1}$. Using the background equations~\eqref{lambda_eq} and \eqref{d_eq}, we find
\begin{align}\label{LBG-redef}
\mathcal{L}_{\rm DE}^{(0)} 
= \frac{\Mpl^2}{2}f \left[ \spatialR +\delta K^{\alpha}{}_{\beta} \delta K^{\beta}{}_{\alpha}-(\delta{K})^2 \right] 
+ \bar{\rho}_{\rm m} - \frac{\bar{N}}{N} \left( \bar{\rho}_{\rm m} + \bar{p}_{\rm m} \right) \,.
\end{align}

\subsection{\texorpdfstring{Quadratic action for the linear perturbations in $\alpha$-basis}{Quadratic action for the linear perturbations in alpha-basis}}
The action~\eqref{action-ST} can be used to study the dynamics of the background, scalar perturbations, and also tensor perturbations at any order of perturbations, while it cannot be used for the vector perturbations since we have ignored the transverse degrees of freedom in the gauge field. In this subsection, we obtain the quadratic action which can be used to study the linear scalar and tensor perturbations.

Substituting \eqref{A-ST} into the action~\eqref{action-ST} and then integrating out $\delta {A}_0$, the quadratic part of $\mathcal{L}_{\rm DE}^{(2)}$ becomes~\cite{Aoki:2021wew}
\begin{align}
\delta_2 \mathcal{L}_{\rm DE}^{(2)} &=\frac{1}{2}
M_{\eff,2}^4(t,k) (\bar{N}^2\delta_1 g^{00})^2
-  \frac{1}{2}\bar{M}_{\eff,1}^3(t,k) (\bar{N}^2\delta_1 g^{00}) \delta_1  K
-  \frac{1}{2}\bar{M}_{\eff,2}^2(t,k) (\delta_1 K)^2
\nn
& \quad
- \frac{1}{2}{\bar M}_3^2(t) \delta_1 K^{\alpha}{}_{\beta} \delta_1 K^{\beta}{}_{\alpha} 
+\frac{1}{2}\mu_{\eff,1}^2(t,k) (\bar{N}^2\delta_1 g^{00}) \delta_1 \! \spatialRST
\nn
& \quad
+\frac{1}{2}\bar{M}_{{\rm eff},4}(t,k) \delta_1 K  \delta_1 \! \spatialRST
+\frac{1}{2}\lambda_{{\rm eff},1}(t,k) (\delta_1 \! \spatialRST)^2 + \cdots \,,
\label{Lquadratic_eff}
\end{align}
with the following $k$-dependent coefficients:
\begin{align}
M_{\eff,2}^4(t,k) &= \frac{k^2/a^2}{\gf^2 M_2^4+k^2/a^2} M_2^4 = [1-\mathcal{G}(t,k)] M_2^4(t)
\,, \label{def_Meff2} \\
\bar{M}_{\eff,1}^3(t,k)&= \frac{k^2/a^2}{\gf^2 M_2^4+k^2/a^2} \bar{M}_1^3 = [1-\mathcal{G}(t,k)] \bar{M}_1^3(t)
\,, \\
\bar{M}_{\eff,2}^2(t,k)&=  \bar{M}_2^2 + \frac{\gf^2 \bar{M}_1^6}{4(\gf^2 M_2^4 +k^2/a^2)} = \bar{M}_2^2(t) + \frac{1}{4}  \frac{\bar{M}_1^6(t)}{M_2^4(t)} \mathcal{G}(t,k)
\,, \\
\mu_{\eff,1}^2(t,k) &= \frac{k^2/a^2}{\gf^2 M_2^4+k^2/a^2} \mu_1^2 = [1-\mathcal{G}(t,k)] \mu_1^2(t)
\,, \\
\bar{M}_{{\rm eff},4}(t,k) &=\bar{M}_4 + \frac{\gf^2 \bar{M}_1^3 \mu_1^2}{2(\gf^2 M_2^4 +k^2/a^2)}= \bar{M}_4(t) +\frac{1}{2}\frac{\bar{M}_1^3(t) \mu_1^2(t)}{M_2^4(t)} \mathcal{G}(t,k)
\,, \\
\lambda_{{\rm eff},1}(t,k) &=\lambda_1-\frac{\gf^2 \mu_1^4}{4(\gf^2 M_2^4 +k^2/a^2)} = \lambda_1(t) -\frac{1}{4} \frac{\mu_1^4(t)}{M_2^4(t)} \mathcal{G}(t,k)
\,,
\label{def_lambda1}
\end{align}
where
\begin{align}\label{g-eff-def}
&\gf(t) \equiv \frac{2 \g \bar{N}}{\sqrt{\gamma_1}(1+\g \bar{A}_0 ) } \,, \qquad
\mathcal{G}(t,k) \equiv \frac{\gf^2 M_2^4}{\gf^2 M_2^4 +k^2/a^2} \,.
\end{align}
Note that the ghost-free condition for vector perturbations requires $\gamma_1>0$. 

The so-called $\alpha$-parameters are defined by~\cite{Bellini:2014fua,Gleyzes:2014qga,Gleyzes:2014rba,Frusciante:2016xoj}
\begin{align}
\begin{split}
&\tilde{\alpha}_B(t,k) \equiv -\frac{\bar{M}_{{\rm eff},1}^3}{2H M^2}
\,, \quad
\alpha_T(t) \equiv \frac{\bar{M}_{3}^2}{M^2}
\,, \quad
\tilde{\alpha}_K(t,k)\equiv  \frac{4M_{{\rm eff},2}^4}{H^2 M^2}
\,, \\
&\tilde{\alpha}_H(t,k) \equiv \frac{2\mu_{{\rm eff},1}^2+\bar{M}_3^2}{M^2}
\,, \quad
\tilde{\alpha}_B^{\rm GLPV}(t,k) \equiv \frac{\bar{M}_3^2+\bar{M}_{{\rm eff},2}^2}{M^2}
\,, 
\\
&\tilde{\alpha}_M^{\rm GC}(t,k) \equiv \frac{\lambda_{{\rm eff},1} H^2}{M^2}
\,, \quad
\tilde{\alpha}_B^{\rm GC}(t,k) \equiv \frac{\bar{M}_{{\rm eff},4} H}{M^2}
\,,
\end{split}
\label{def_alpha}
\end{align}
where
\begin{align}
M^2(t) \equiv \Mpl^2 f - \bar{M}_3^2
\,, \label{def_M}
\end{align}
is the effective Planck mass for the tensor perturbations. Defining 
\begin{align}\label{alpha-V}
\alpha_V \equiv \frac{ \g^2 \bar{N}^2 M^2}{\gamma_1(1+\g \bar{A}_0 )^2 } \,,
\end{align}
and then using the definition~\eqref{eq:def_G}, we find \eqref{alpha_relation}. Note that $\alpha_V=0$ corresponds to $\g=0$. Therefore, the parameter~$\alpha_V$ determines the boundary between the scalar-tensor and vector-tensor theories.

Using the expression~\eqref{LBG-redef} and also the functions defined in \eqref{def_alpha} or equivalently \eqref{alpha_relation}, it is straightforward to show that the quadratic action for the scalar and tensor perturbations in the momentum space is given by
\begin{align}
S_2 = \int &\frac{\D t \D^3k}{(2\pi)^3} \bar{N}a^3  \frac{M^2}{2}
\nn
\times \Biggl[ & 
(1+\tilde{\alpha}_H) \frac{\delta_1 N}{\bar{N}} \delta_1 \! \spatialR + 4 H \tilde{\alpha}_B \frac{\delta_1 N}{\bar{N}} \delta_1 K
+\delta_1 K^{\alpha}{}_{\beta} \delta_1 K^{\beta}{}_{\alpha}
-(1+\tilde{\alpha}_B^{\rm GLPV}) (\delta_1 K)^2 
\nn
&
+ \tilde{\alpha}_K H^2 \left(\frac{\delta_1 N}{\bar{N}}\right)^2
+(1+\alpha_T) \delta_2\bigg( \spatialR \frac{\sqrt{h}}{a^3}\, \bigg)  + \frac{\tilde{\alpha}_M^{\rm GC}}{H^2} (\delta_1 \! \spatialR)^2
+ \frac{\tilde{\alpha}_B^{\rm GC} }{H} \delta_1 K  \delta_1 \! \spatialR + \cdots \Biggl] 
+ S^{\rm m}_2 \,,
\label{LEFT_alpha-app}
\end{align}
where 
\begin{align}
S^{\rm m}_2 \equiv \delta_2 S^{\rm m} 
+ \delta_2 \int \D^4 x \bar{N} \sqrt{h} \left( \bar{\rho}_{\rm m} \frac{\delta{N}}{\bar N} - \bar{p}_{\rm m} \right) \,.
\end{align}
The above form of the quadratic action is used in this paper to study the linear scalar perturbations in the unified EFT of DE.


\bibliographystyle{JHEPmod}
\bibliography{refs}

\end{document}